\documentclass[12pt,a4paper]{article}
\usepackage[dvips]{graphicx}
\usepackage{amsmath}
\usepackage{amssymb}
\usepackage{cite}
\newcommand{\rd}{{\rm d}}
\newcommand{\re}{{\rm e}}
\newcommand{\ri}{{\rm i}}

\newcommand{\stat}{s}
\renewcommand{\textrm}[1]{\mbox{#1}}

\DeclareGraphicsExtensions{.eps}

\begin{document}

\title{On two-dimensional Bessel functions}

\author{H. J. Korsch, A. Klumpp\footnote{Present address: Institut f\"ur Physik,
Universit\"at Kassel, D-34109 Kassel, Germany}, and D. Witthaut\\
FB Physik, Technische Universit\"at Kaiserslautern\\ D-67653 Kaiserslautern, Germany}

\maketitle

\begin{abstract}
The general properties of two-dimensional generalized
Bessel functions are discussed. Various asymptotic approximations
are derived and applied to analyze the basic structure of the
two-dimensional Bessel functions as well as their nodal lines.
\end{abstract}

\section{Introduction}
\label{s-intro}

Generalized Bessel functions depending on several variables have been
introduced
in 1915 for a finite \cite{Appe15} and also an infinite number of
variables \cite{Pere15}. They have very similar properties as the
ordinary Bessel functions but are much less familiar. 
More recently, however,
they found an increasing number of applications in various areas
of physics (see, e.g.,
\cite{Niki64,Wasi69,Reis80,Beck86,Beck87,Datt92,Paci94,Reis03,02tb2d,03TBalg,Baue05}). 
The basic theory of generalized Bessel functions
is described in a mongraph by Dattoli and Torre \cite{Datt96}.
Our own interest into the properties of these functions is caused
by our recent studies of quantum dynamics in periodic structures
\cite{02tb2d,03TBalg},
in particular in studies of transport and dynamic localization
\cite{05bicro}.

In most cases these applications were restricted to the case
of two variables, $u$ and $v$. Then the generalized Bessel functions 
\,$J_{n}^{p,q}(u,v)$\, are labeled by three integer indices $n,p,q$.
The special case $(p,q)=(1,2)$ has been considered up to now
almost exclusively
\cite{Reis80,Reis03,Datt90,Datt91a,Datt91b,Datt92,Datt96}.  Here we will analyze
the two-dimensional Bessel functions \,$J_{n}^{p,q}(u,v)$\, 
for general indices $p$ and $q$ (see \cite{Lore95} for
a well written introduction to the case of infinite variables).

We will derive the fundamental properties of the
two-dimensional Bessel functions and analyze their basic 
structure for small and large arguments
in the following sections.
It will be seen that the two-dimensional Bessel functions show a rich
oscillatory structure with regions of very different behavior. We
will analyze these structural features with special attention to
the nodal lines which are of considerable importance for recent
applications to localization phenomena in quantum dynamics 
\cite{05bicro}.
\section{Basic properties}
\label{s-basic}
In this section we will collect the basic properties of the 
generalized Bessel functions
\,$J_{n}^{p,q}(u,v)$\, with integer indices $n,\,p,\,q$ 
and two real arguments $u,\,v$. Most of the results
in the literature (see, in particular, appendix B of \cite{Reis80} and chapter 2 of
\cite{Datt96}) have been 
derived for the special case of \,$J_{n}^{1,2}(u,v)$\,. 

\subsection{Definition}
The two-dimensional Bessel functions can be defined by the
generating function 
\begin{eqnarray}
\re^{\ri (u\sin pt+v\sin qt)}=
\sum\limits_{n=-\infty}^\infty J_{n}^{p,q}(u,v)\,\re^{\ri nt}\,,
\label{gen-funct}
\end{eqnarray}
also known as a Jacobi-Anger expansion, or, somewhat more general, as
\begin{eqnarray}
\exp \Big(\,\frac{u}{2}(z^p-z^{-p})+ \frac{v}{2}(z^q-z^{-q})\,\Big)=
\sum\limits_{n=-\infty}^\infty J_{n}^{p,q}(u,v)\,z^n\,.
\label{gen-funct-gen}
\end{eqnarray}
Integration of (\ref{gen-funct}) over $t$ using 
\,$\int_{-\pi}^{+\pi}\rd t\,\re^{\ri nt}=2\pi\delta_{n,0}$ immediately
leads to the integral representation
\begin{eqnarray}
J_{n}^{p,q}(u,v)=\frac{1}{2\pi}
\int_{-\pi}^{+\pi}\rd t\,
\re^{\ri (u\sin pt+v\sin qt-nt)}\,,
\label{int-rep}
\end{eqnarray}
a generalization of the integral representation of the
well known ordinary Bessel function
\begin{eqnarray}
J_{n}(x)=\frac{1}{2\pi}
\int_{-\pi}^{+\pi}\rd t\,
\re^{\ri (x\sin t-nt)}\,.
\label{int-rep-ord-bessel}
\end{eqnarray}
From the properties of Fourier series we immediately find the bounds
\begin{eqnarray}
|J_{0}^{p,q}(u,v)|\le 1 \ \textrm{ and } \ |J_{n}^{p,q}(u,v)|\le 1/\sqrt{2}
\ \textrm{  for  } \ n\ne 0\,.
\label{bounds}
\end{eqnarray}
As an immediate consequence of (\ref{int-rep})
the integers $p$ and $q$ can be assumed to be coprime because
$J_n^{p,q}$ vanishes otherwise or it can be reduced to such a
coprime case.
 This is seen as follows (we assume that
$\mu$ is an integer, $0\ne \mu \ne 1$):
\begin{eqnarray}
&&2\pi J_n^{\mu p,\mu q}(u,v)
=\int_{0}^{2\pi}\rd t\,\re^{\ri (u\sin \mu pt+v\sin \mu qt-nt)}\nonumber \\
&&=\int_{0}^{2\pi\mu}\frac{\rd s}{\mu}\,\re^{\ri (u\sin ps+v\sin qs-ns/\mu )}
=\sum_{m=1}^\mu \int_{2\pi (m-1)}^{2\pi m}\frac{\rd s}{\mu}\,\re^{\ri (u\sin ps+v\sin qs-ns/\mu )}\nonumber\\
&&=\Big[\sum_{m=1}^\mu \re^{-\ri 2\pi (m-1)n/\mu}\Big]
\int_{0}^{2\pi}\frac{\rd s}{\mu}\,\re^{\ri (u\sin ps+v\sin qs-ns/\mu )}\,.
\end{eqnarray}
Using
\begin{eqnarray}
\sum_{m=1}^\mu \re^{-\ri 2\pi (m-1)n/\mu}=
\left\{\begin{array}{ll}
\mu \quad & \textrm{for  } n/\mu \in \mathbb{Z}\\
0 & \textrm{else}\end{array}\right.
\end{eqnarray}
and -- for $n/\mu \in \mathbb{Z}$ --
\begin{eqnarray}
\int_{0}^{2\pi}\frac{\rd s}{\mu}\,\re^{\ri (u\sin ps+v\sin qs-ns/\mu )}
=\frac{1}{\mu}\,J_{n/\mu}^{p,q}(u,v)
\end{eqnarray}
one obtains
\begin{eqnarray}
J_n^{\mu p,\mu q}(u,v)=
\left\{\begin{array}{ll}
J_{n/\mu}^{p,q}(u,v) \quad & \textrm{for  } n/\mu \in \mathbb{Z}\\
0 & \textrm{else}\,.\end{array}\right.
\end{eqnarray}
In the following, we will therefore assume that the integers $p$ and $q$ have no 
common divisor.
\subsection{Decomposition in terms of ordinary Bessel functions}
A representation in terms of ordinary Bessel functions can be
derived from the integral representation
(\ref{int-rep}). 
Inserting the generating function for the ordinary
Bessel functions
\begin{eqnarray}
\re^{\ri x\sin s}=
\sum\limits_{n=-\infty}^\infty J_{n}(x)\,\re^{\ri ns}
\label{gen-funct-ob}
\end{eqnarray}
for both $s=pt$ and $s=qt$ into (\ref{int-rep}), we obtain
\begin{eqnarray}
J_{n}^{p,q}(u,v)&=&\frac{1}{2\pi}
\int_{-\pi}^{+\pi}\rd t\,
\re^{\ri (u\sin pt+v\sin qt-nt)}\nonumber\\
&=&\sum_{\mu,\nu } J_\mu (u)\,J_\nu(v)
\frac{1}{2\pi}\int_{-\pi}^{+\pi}\rd t\,
\re^{\ri [(\mu p+\nu q-n)t]}
\,.
\end{eqnarray}
The integral is only different from zero if \,$n=\mu p+\nu q$\,
is satisfied. If $p$ and $q$ have no common divisor as assumed here,
a solution  $(\mu,\nu)=(M,N)$ of this Diophantine equation always exists and
can be found systematically by, e.g., the Euclid algorithm \cite{Mord69}.
Moreover there is
an infinite number of solutions \,$\mu =M-qk$, $\nu=N+pk$,
$k=0,\pm 1,\pm 2,\,...$\, because of
\begin{eqnarray}\label{Dioph}
\nonumber
n=pM+qN
=pM+qN+pqk-pqk
=p(M-qk)+q(N+pk)\,.
\end{eqnarray}
We therefore have
\begin{eqnarray}
J_{n}^{p,q}(u,v)&=&\sum\limits_{k=-\infty}^\infty J_{M-qk}(u)\, J_{N+pk}(v)\,,
\label{gen-bessel-sum}
\end{eqnarray}
where $(M,N)$ is an arbitrary solution of $n=pM+qN$. 

For the case $p=1$  this reads ($M=n$ and $N=0$)
\begin{eqnarray}
J_{n}^{1,q}(u,v)=\sum\limits_{k=-\infty}^\infty J_{n-qk}(u)\, J_{k}(v)
\,.
\label{gen-bessel-sum-1q}
\end{eqnarray}
In most of the previous applications one encounters the case $q=2$ and
in these cases one usually simplifies the notation by dropping the
$p,q$ indices, i.e.~one defines 
\begin{eqnarray}
J_n(u,v)=J_n^{1,2}(u,v)\,.
\label{gen-bessel-sum-12}
\end{eqnarray}
\subsection{Addition theorems}
The addition theorem
\begin{eqnarray}
J_{n}^{p,q}(u_1+u_2,v_1+v_2)
=\sum\limits_{k=-\infty}^\infty J_{n-k}^{p,q}(u_1,v_1)\,J_{k}^{p,q}(u_2,v_2)
\label{add-theorem}
\end{eqnarray}
can be easily proved starting from (\ref{int-rep}) 
using (\ref{gen-funct}):
\begin{eqnarray}
&&J_{n}^{p,q}(u_1+u_2,v_1+v_2)=\frac{1}{2\pi}
\int_{-\pi}^{+\pi}\rd t\,
\re^{\ri (u_1\sin pt+v_1\sin qt-nt)}\,
\re^{\ri (u_2\sin pt+v_2\sin qt)}\nonumber\\
&&=\frac{1}{2\pi}
\int_{-\pi}^{+\pi}\rd t\,
\re^{\ri (u_1\sin pt+v_1\sin qt-nt)}
\sum\limits_{k =-\infty}^\infty J_{k}^{p,q}(u_2,v_2)\,\re^{\ri k t}\nonumber\\
&&=  \sum\limits_{k =-\infty}^\infty J_{k}^{p,q}(u_2,v_2)\ \frac{1}{2\pi}
\int_{-\pi}^{+\pi}\rd t
\,\re^{\ri (u_1\sin pt+v_1\sin qt-(n-k )t)}\nonumber\\
&&=\sum\limits_{k =-\infty}^\infty J_{k}^{p,q}(u_2,v_2)\,
J_{n-k}^{p,q}(u_1,v_1)\,.\nonumber
\end{eqnarray}
The Graf addition theorem for ordinary Bessel functions,
\begin{eqnarray}
\sum_{\ell=-\infty}^{+\infty}\tau^\ell J_\ell(x_1)J_{n+\ell}(x_2)
=\left[\frac{x_2-x_1/\tau}{x_2-x_1\tau}\right]^{\frac{n}{2}}J_n[g(x_1,x_2;\tau)] 
\label{graf}
\end{eqnarray}
with
\begin{eqnarray}
\qquad g(x_1,x_2;\tau)=\big(x_1^2+x_2^2-x_1x_2(\tau+1/\tau)\big)^{1/2}\,,
\label{graf-g}
\end{eqnarray}
can also be generalized to the two-dimensional case, at least for
$p=1$ in the form
\begin{eqnarray}
&&\sum_{\ell=-\infty}^{+\infty}\tau^\ell J_\ell^{1,q}(u_1,v_1)J_{n+\ell}^{1,q}(u_2,v_2)
\label{graf-gen}\\[2ex]
&&=\sum_{\ell=-\infty}^{+\infty}
\left[\frac{u_2-u_1/\tau}{u_2-u_1\tau}\right]^{\frac{n-q\ell}{2}}
\left[\frac{v_2-v_1/\tau^q}{v_2-v_1\tau^q}\right]^{\frac{\ell}{2}}
J_{n-q\ell}[g(u_1,u_2;\tau)]\,J_{\ell}[g(v_1,v_2;\tau^q)] 
\nonumber
\end{eqnarray}
(see Dattoli et al. \cite{Datt90,Datt96} for the special case $q=2$).
The generalized Graf addition theorem (\ref{graf-gen}) can be derived
in a straightforward calculation expressing first the  two-dimensional Bessel functions
as a sum over ordinary ones (see eqn.~(\ref{gen-bessel-sum-1q})) and using
the Graf addition theorem (\ref{graf}) for ordinary Bessel functions:
\begin{eqnarray}
&&\sum_{\ell=-\infty}^{+\infty}\tau^\ell J_\ell^{1,q}(u_1,v_1)J_{n+\ell}^{1,q}(u_2,v_2)\\
&&\ =\sum_{\ell, j,k}\tau^\ell J_{\ell-qk}(u_1)J_k(v_1)J_{\ell+n-qj}(u_2)J_j(v_2)\nonumber\\
&&\ =\sum_{j,k} J_k(v_1)J_j(v_2)\,\tau^{qk}\sum_{\ell'}\tau^{\ell'} J_{\ell'}(u_1)J_{n+q(k-j)+\ell'}(u_2)\nonumber\\
&&\ =\sum_{j,k} J_k(v_1)J_j(v_2)\,\tau^{qk}J_{n+q(k-j)}(g(u_1,u_2;\tau))
\left[\frac{u_2-u_1/\tau}{u_2-u_1\tau}\right]^{\frac{n+q(k-j)}{2}}\nonumber\\
&&\ =\sum_{\ell} \left[\frac{u_2-u_1/\tau}{u_2-u_1\tau}\right]^{\frac{n-q\ell}{2}}
\!\!J_{n-q\ell}(g(u_1,u_2;\tau))\sum_{k}\tau^{qk}J_k(v_1)J_{\ell+k}(v_2)\nonumber\\
&&\ =\sum_{\ell} \left[\frac{u_2-u_1/\tau}{u_2-u_1\tau}\right]^{\frac{n-q\ell}{2}}
\!\!\left[\frac{v_2-v_1/\tau^q}{v_2-v_1\tau^q}\right]^{\frac{\ell}{2}}
\!\!J_{n-q\ell}(g(u_1,u_2;\tau))J_\ell(g(v_1,v_2;\tau^q))\nonumber
\end{eqnarray}
with \,$g(x_1,x_2;\tau)$\, as defined 
in (\ref{graf-g}). 
\subsection{Symmetries, special cases and numerical examples}
From the definition (\ref{gen-funct})
one verifies (by taking
the complex conjugate and changing variables $t\rightarrow -t$) that the
\,$J_{n}^{p,q}(u,v)$\, are real valued and satisfy
\begin{eqnarray}
J_{n}^{p,q}(0,0)=\delta_{n 0}\,.
\label{uv-zero}
\end{eqnarray}
The symmetry relations
\begin{eqnarray}
J_{n}^{p,q}(u,v)&=&J_{n}^{q,p}(v,u)\\
J_{-n}^{p,q}(u,v)&=&J_{n}^{p,q}(-u,-v)=J_{n}^{q,p}(-v,-u)=J_{n}^{-p,-q}(u,v)
\,.
\label{properties1}
\end{eqnarray}
follow directly from the definition.
For $n=0$ these equations imply the symmetries
\begin{eqnarray}
J_{0}^{p,q}(u,v)=J_{0}^{p,q}(-u,-v)=J_{0}^{q,p}(v,u)\,.
\label{symm-n-0}
\end{eqnarray}

A further direct result is a symmetry relation for even values of one of the $p,q$-indices,
say $q$. Using $J_n(-z)=(-1)^n\,J_n(z)$ we get 
\begin{eqnarray}
&&J_{n}^{p,q}(-u,v)= \sum\limits_{k=-\infty}^\infty J_{M-qk}(-u)\, J_{N+pk}(v)
\nonumber\\
&&\quad=(-1)^M\sum\limits_{k=-\infty}^\infty J_{M-qk}(u)\, J_{N+pk}(v)
=(-1)^n\,J_n^{p,q}(u,v)\,.
\label{symm-q-even}
\end{eqnarray}
The last equality holds because of $(-1)^n=(-1)^{pM+qN}=(-1)^M$ for $q$ even and $p$ odd. 
This symmetry implies
\begin{eqnarray}
J_{n}^{p,q}(0,v)=(-1)^nJ_{n}^{p,q}(0,v)
\ \Longrightarrow \ J_{n}^{p,q}(0,v)=0 \textrm{  for  } n \textrm{  odd and  }
q \textrm{ even}.
\label{symm-u=0}
\end{eqnarray}
If both upper indices are odd, their difference must be even. This leads to
another symmetry relation
\begin{eqnarray}
&&J_{n}^{p,q}(-u,-v)= \sum\limits_{k=-\infty}^\infty J_{M-qk}(-u)\, J_{N+pk}(-v)
\nonumber\\
&&\qquad =\sum\limits_{k=-\infty}^\infty (-1)^{M+N+(p-q)k}J_{M-qk}(u)\, J_{N+pk}(v)\\[1mm]
&&\qquad =(-1)^{M+N}\,J_n^{p,q}(u,v)=(-1)^n\,J_n^{p,q}(u,v)
\quad \textrm{for } p,q \textrm{ odd}\,.\nonumber
\label{symm-pq-odd}
\end{eqnarray}
Here the last equality is based on the fact that for odd $p,q$-indices, $p=2j+1$ 
and $q=2k+1$, we have $n=pM+qN=M+N+2(j+k)$. 

In the case $p=q$ the two-dimensional Bessel functions simplify
and reduce to ordinary Bessel functions if $n$ is an integer multiple of $p$:
\begin{eqnarray}
J_{n}^{p,p}(u,v)&=&\frac{1}{2\pi}
\int_{-\pi}^{+\pi}\rd t\,
\re^{\ri ((u+v)\sin pt-nt)}
=\frac{1}{2\pi}\int_{-p\pi}^{+p\pi}\frac{\rd s}{p}\,
\re^{\ri ((u+v)\sin s-ns/p)}\nonumber\\[2mm]
&=&\left\{\begin{array}{ll}
J_{n/p}(u+v) \quad & \textrm{for  } n/p \in \mathbb{N}\\
0 & \textrm{else}\,.\end{array}\right.
\label{int-rep-p=q}
\end{eqnarray}
Another relation between the generalized and ordinary Bessel
functions can be observed if the index $n$ is a multiple of one
of the upper indices, e.g. $n=mq$, $m$ integer. Then we get
\begin{eqnarray}
J_{mq}^{p,q}(0,v)&=&
\frac{1}{2\pi}
\int_{-\pi}^{+\pi}\rd t\,\re^{\ri (v\sin qt-mqt)}
=\frac{1}{2\pi q}
\int_{-q\pi}^{q\pi}\rd s\,\re^{\ri (v\sin s-ms)}\nonumber\\
&=&\frac{1}{2\pi}
\int_{-\pi}^{+\pi}\rd s\,\re^{\ri (v\sin s-ms)}=J_m(v)
\label{u=0-n=mq}
\end{eqnarray}
and, as a special case, 
\begin{eqnarray}
J_{n}^{p,1}(0,v)=J_{n}(v)\,.
\label{pq=1q-v=0}
\end{eqnarray}

According to (\ref{u=0-n=mq}) the two-dimensional Bessel function 
$J_n^{p,q}(0,v)$ is
reduced to an ordinary Bessel function $J_m(v)$ for $n=mq$.
Otherwise the function
vanishes on the  $v$-axis as can easily be seen from (\ref{gen-bessel-sum}):
\begin{eqnarray}
J_{n}^{p,q}(0,v)\!=\!\sum\limits_{k=-\infty}^\infty J_{M-qk}(0)\, J_{N+pk}(v)
\!=\!\sum\limits_{k=-\infty}^\infty \delta_{M,qk}\, J_{N+pk}(v)=0\,,
\label{symm-u0}
\end{eqnarray}
if $M$ is not a multiple of $q$ or, equivalently, \,$n=pM+qN$\, is not an
integer multiple of $q$. We therefore have 
\begin{eqnarray}
J_{n}^{p,q}(0,v)=0\quad \textrm{if}\ n\ne mq,\ m\in
\mathbb{Z}
\label{symm-n=mq}
\end{eqnarray}
as a generalization of  (\ref{symm-u=0}).

\begin{figure}[t]
\center
\includegraphics[width=4.4cm,  angle=0]{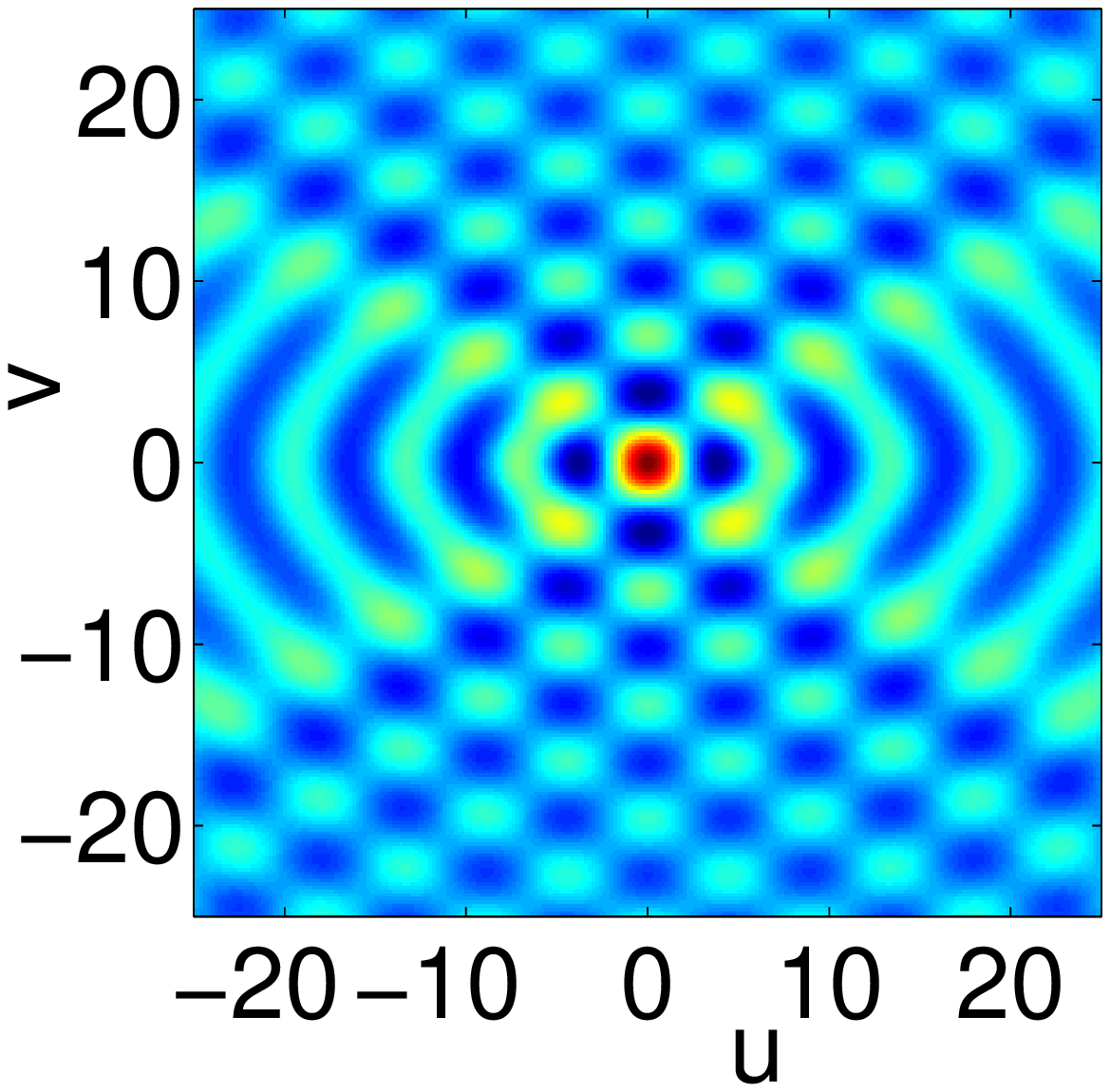}
\includegraphics[width=4.4cm,  angle=0]{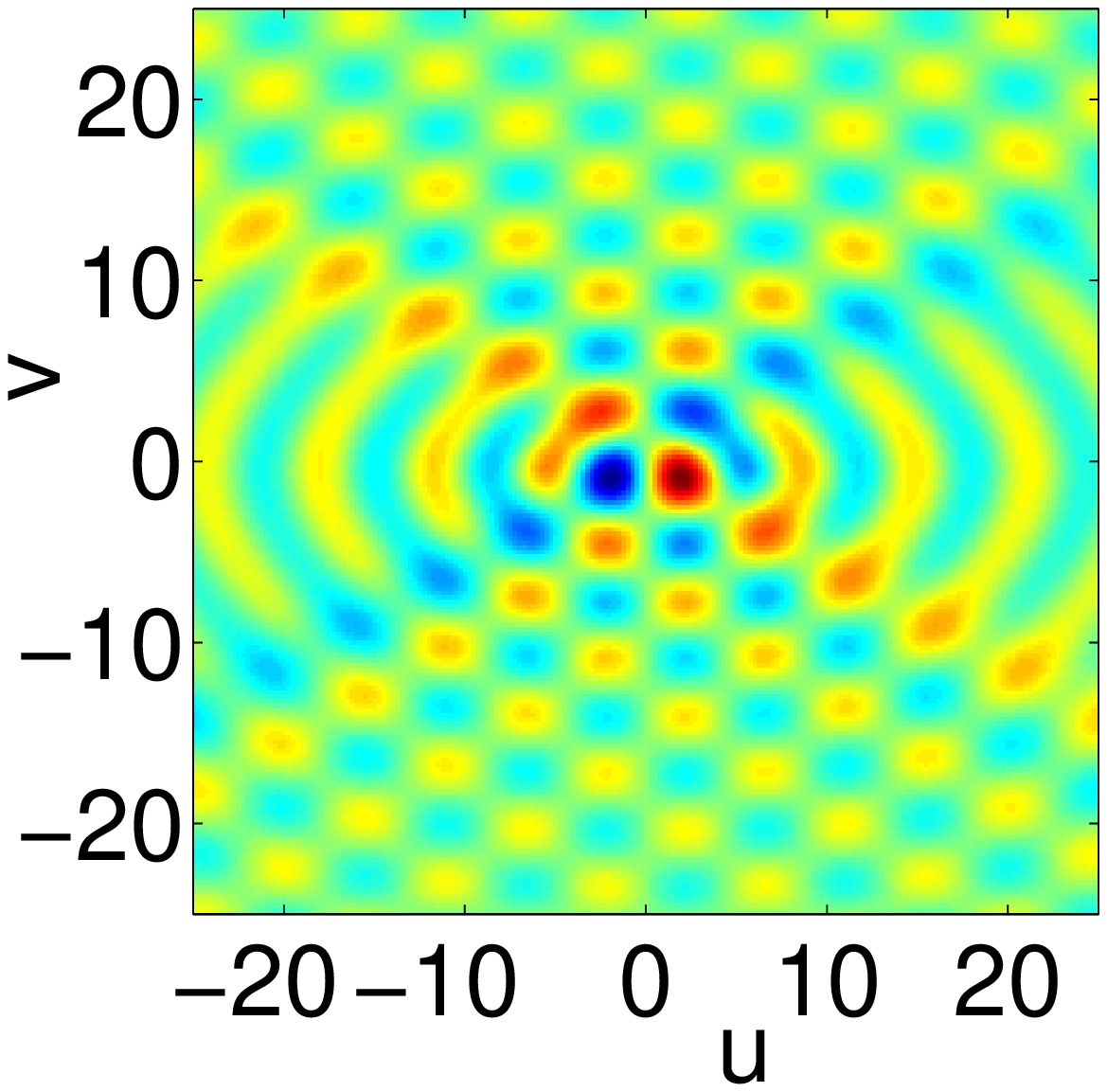}
\includegraphics[width=4.4cm,  angle=0]{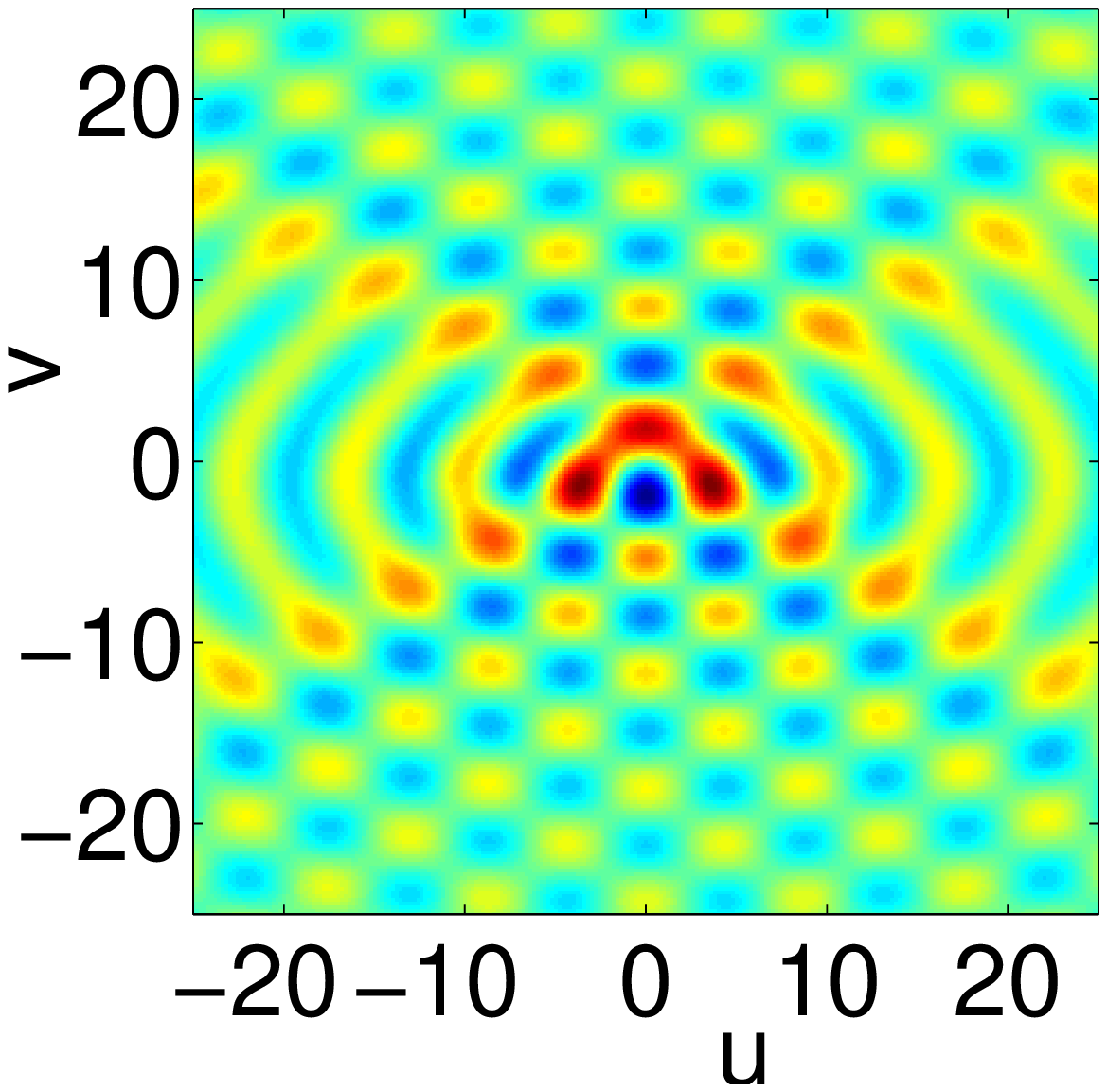}
\caption{\label{bessel_12_n}
Color map of the two-dimensional Bessel function $J_n^{1,2}(u,v)$ for $n=0$, $n=1$, $n=2$ (from
left to right).}
\end{figure}
\begin{figure}[t]
\center
\includegraphics[width=4.4cm,  angle=0]{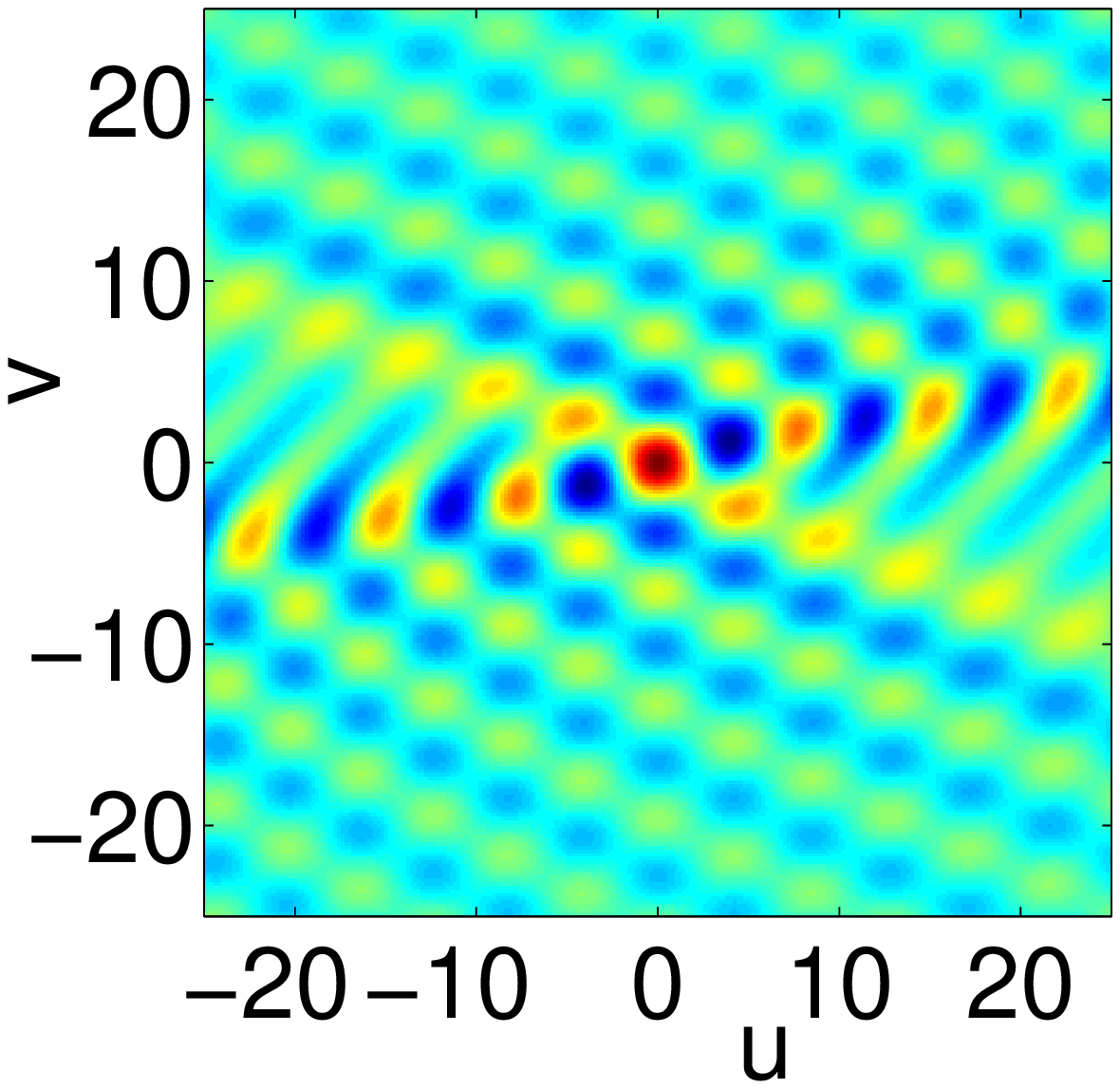}
\includegraphics[width=4.4cm,  angle=0]{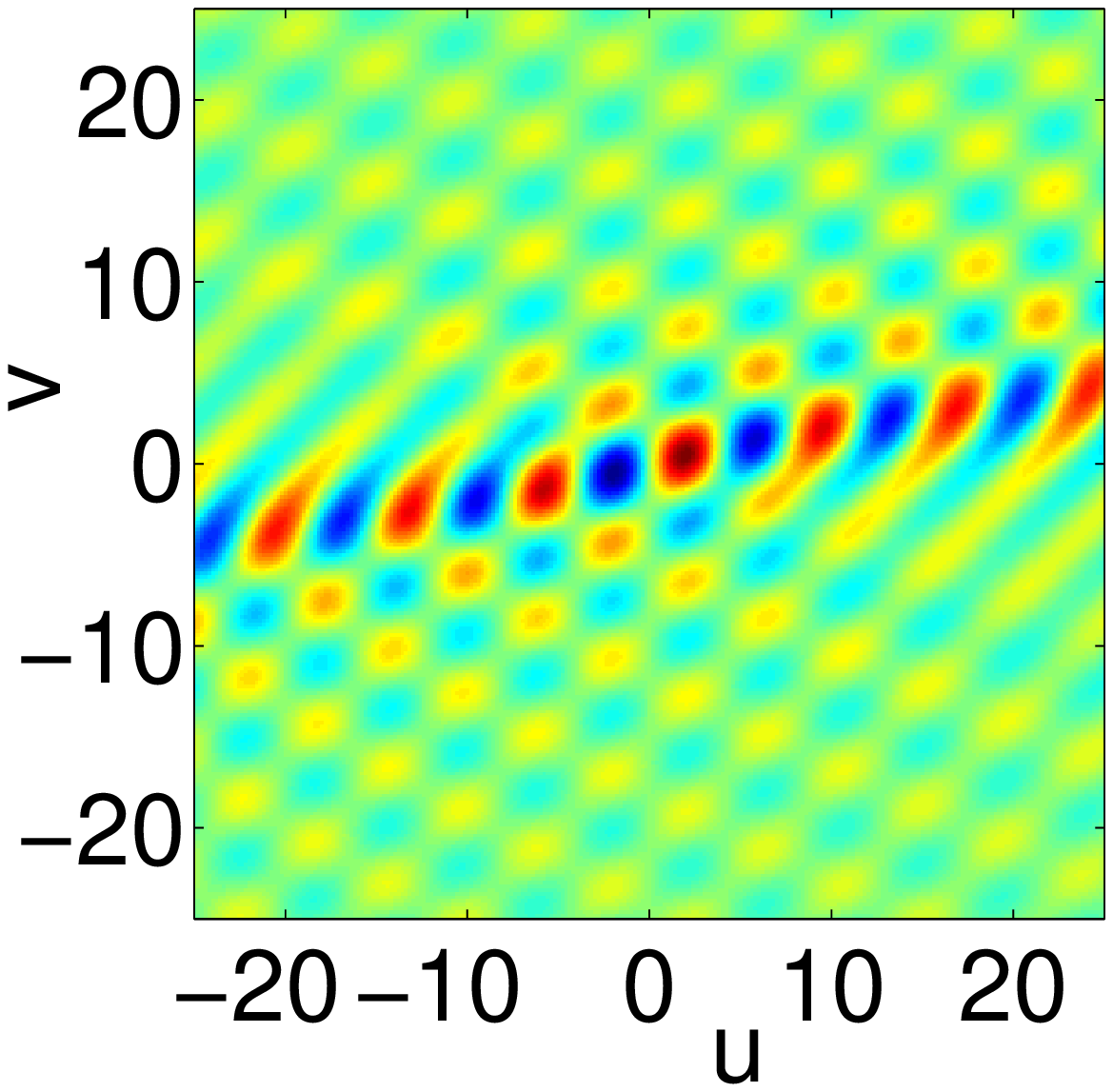}
\includegraphics[width=4.4cm,  angle=0]{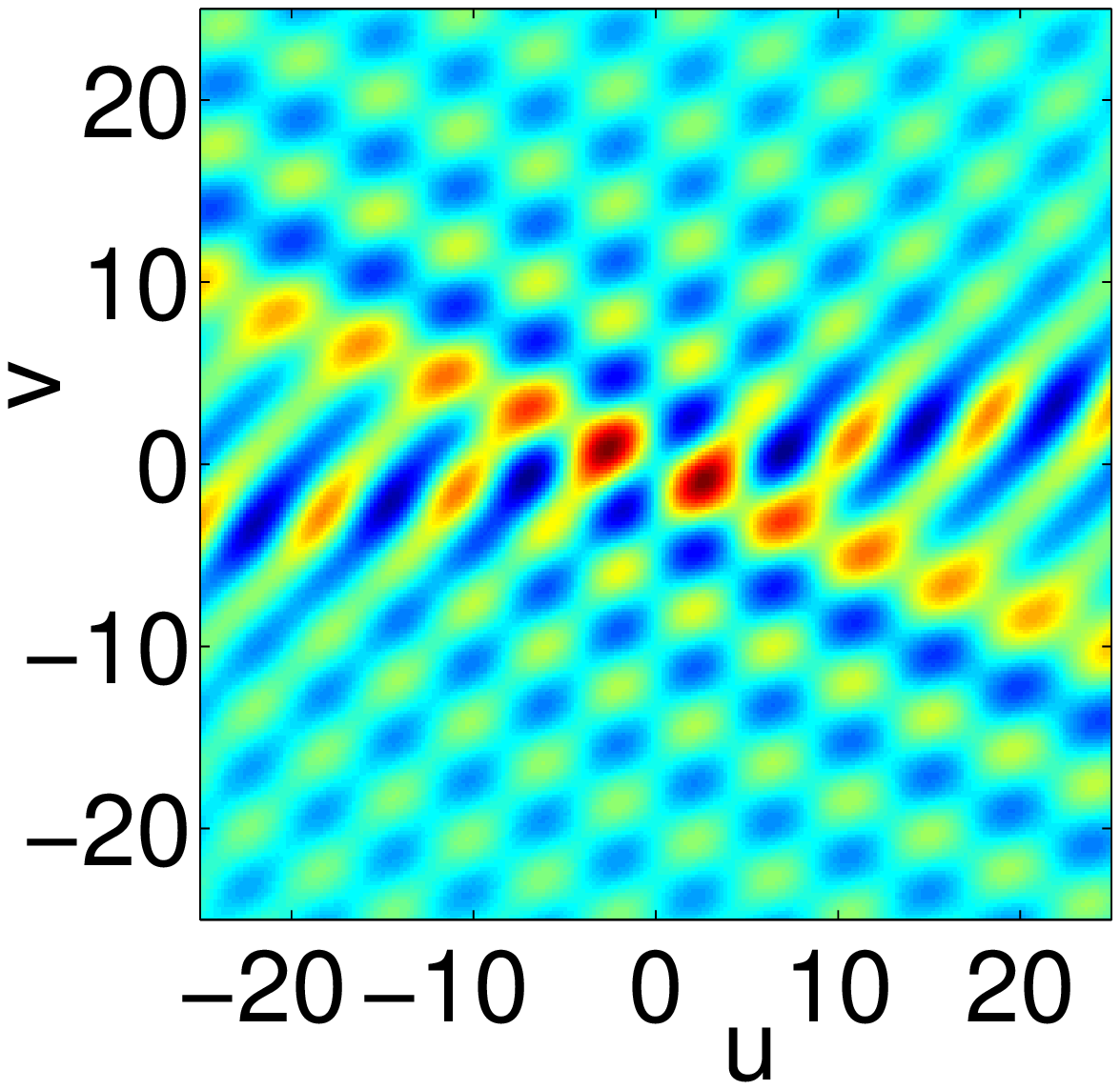}
\caption{\label{bessel_13_n}
Color map of the two-dimensional Bessel function $J_n^{1,3}(u,v)$ for $n=0$, $n=1$, $n=2$ (from
left to right).}
\end{figure}
Let us look at a few examples of two-dimensional Bessel functions calculated
numerically using the representation (\ref{gen-bessel-sum}) in terms of
ordinary Bessel functions (similar graphs can be found, e.g., in \cite{Datt91a,Datt96}).
Figures \ref{bessel_12_n} and  \ref{bessel_13_n} show color maps of 
$J_n^{1,2}(u,v)$ and $J_n^{1,3}(u,v)$  for $n=0$, $n=1$ and $n=2$ using
a re-normalization to unit maximum in each case (the regions of 
positive values
are colored red, of negative ones blue).

For $J_{n}^{1,2}$  in  figure \ref{bessel_12_n} we see that the symmetry relations 
\begin{eqnarray}
J_{0}^{1,2}(-u,-v)=J_{0}^{1,2}(u,v) \quad ,\quad
J_n^{1,2}(-u,v)=(-1)^n\,J_n^{1,2}(u,v)
\label{symm12}
\end{eqnarray}
are satisfied
(compare eqs.~(\ref{symm-n-0}) and (\ref{symm-q-even})\,). The last relation 
implies a nodal line for $n=1$ along the $v$-axis,
\,$J_1^{1,2}(0,v)=0$.
For the case  $J_{n}^{1,3}(-u,-v)$  in  figure \ref{bessel_13_n} we have the 
symmetry \,$J_n^{1,3}(-u,-v)=(-1)^n\,J_n^{1,3}(u,v)$\,
(compare eq.~(\ref{symm-pq-odd})).

\subsection{Sum rules and Kapteyn series}
The simple sum rule
\begin{eqnarray}
\sum\limits_{n=-\infty}^\infty J_{n}^{p,q}(u,v)=1
\label{sum-rule-1}
\end{eqnarray}
is a direct consequence of the generating function (\ref{gen-funct}) for $t=0$.
A variety of sum rules for special cases can be obtained by choosing
$t$ in (\ref{gen-funct}) appropriately. E.g.~for the important special case 
$(p,q)=(1,2)$ another sum rule is found by setting $t=\pi/2$:
\begin{eqnarray}
\sum\limits_{n=-\infty}^\infty \ri^n\,J_{n}^{1,2}(u,v)=\re^{\ri u}\,.
\end{eqnarray}
Similar sum rules can be obtained for other special cases.
Another sum rule, 
\begin{eqnarray}
\sum\limits_{k=-\infty}^\infty \big(J_{k}^{p,q}(u,v)\big)^2=1
\label{sum-rule-2}
\end{eqnarray}
follows from the addition theorem (\ref{add-theorem})
for the special case $n=0$, $u_1=-u_2=u$,  $v_1=-v_2=v$ using
(\ref{uv-zero}). 

We furthermore note without proof the Kapteyn type series \cite{Datt98,Datt04}
\begin{eqnarray}
\sum\limits_{n=-\infty}^\infty J_{n}^{p,q}(nu,nv)=\frac{1}{1-pu-qv}\quad ,\quad |pu|+|qv|<1\,.
\end{eqnarray}

\subsection{Further generalizations}
As already stated in the introduction, the number of variables in the Bessel function
can be extended. Different types of generalizations are, however, also possible. 
Modified higher dimensional Bessel functions can be constructed, e.g.~by replacing
one of the ordinary Bessel functions in (\ref{gen-bessel-sum}) by a modified one
\cite{Paci94}. In addition, two-variable, one-parameter Bessel functions
\cite{Paci94,Datt02} can be defined as a generalization of (\ref{gen-bessel-sum}):
\begin{eqnarray}
J_{n}^{p,q}(u,v;\tau)&=&\sum\limits_{k=-\infty}^\infty J_{M-qk}(u)\, J_{N+pk}(v) \,\tau^k 
\label{genX-bessel-sum}
\end{eqnarray}
(Let us recall that $(M,N)$ are arbitrary solutions of  $n=pM+qN$.) Here again 
we find \,$J_{n}^{p,q}(0,0;\tau)=\delta_{n 0}$\,.

In particular the case $\tau = \re^{\ri \delta}$ is of interest \cite{Paci94,05bicro}
for applications in physics. We confine ourselves to the most important
case $p=1$, i.e.
\begin{eqnarray}
J_{n}^{1,q}(u,v;\re^{\ri \delta} )&=&\sum\limits_{k=-\infty}^\infty J_{n-qk}(u)\, J_{k}(v) 
 \re^{\ri k\delta}\,.
\label{genp-bessel-sum-1q}
\end{eqnarray}
Following the lines in the derivations above, one can easily show 
that these functions are generated by
\begin{eqnarray}
\re^{\ri (u\sin t+v\sin (qt+\delta))}=
\sum\limits_{n=-\infty}^\infty J_{n}^{1,q}(u,v; \re^{\ri \delta})\,\re^{\ri nt}\,,
\label{genp-funct}
\end{eqnarray}
which leads to the integral representation
\begin{eqnarray}
J_{n}^{1,q}(u,v;\re^{\ri \delta})=\frac{1}{2\pi}
\int_{-\pi}^{+\pi}\rd t\,
\re^{\ri (u\sin t+v\sin (qt+\delta)-nt)}\,.
\label{p-int-rep}
\end{eqnarray}
The generalized Bessel functions \,$J_{n}^{1,q}(u,v;\re^{\ri \delta})$\,
satisfy most of the properties of the  \,$J_{n}^{1,q}(u,v)$\,,
as for example the bounds (\ref{bounds}), the
addition theorem (\ref{add-theorem}) and the sum rules
(\ref{sum-rule-1}) and (\ref{sum-rule-2}). These function are, however, complex valued.

\begin{figure}[t]
\center
\includegraphics[width=4.4cm,  angle=0]{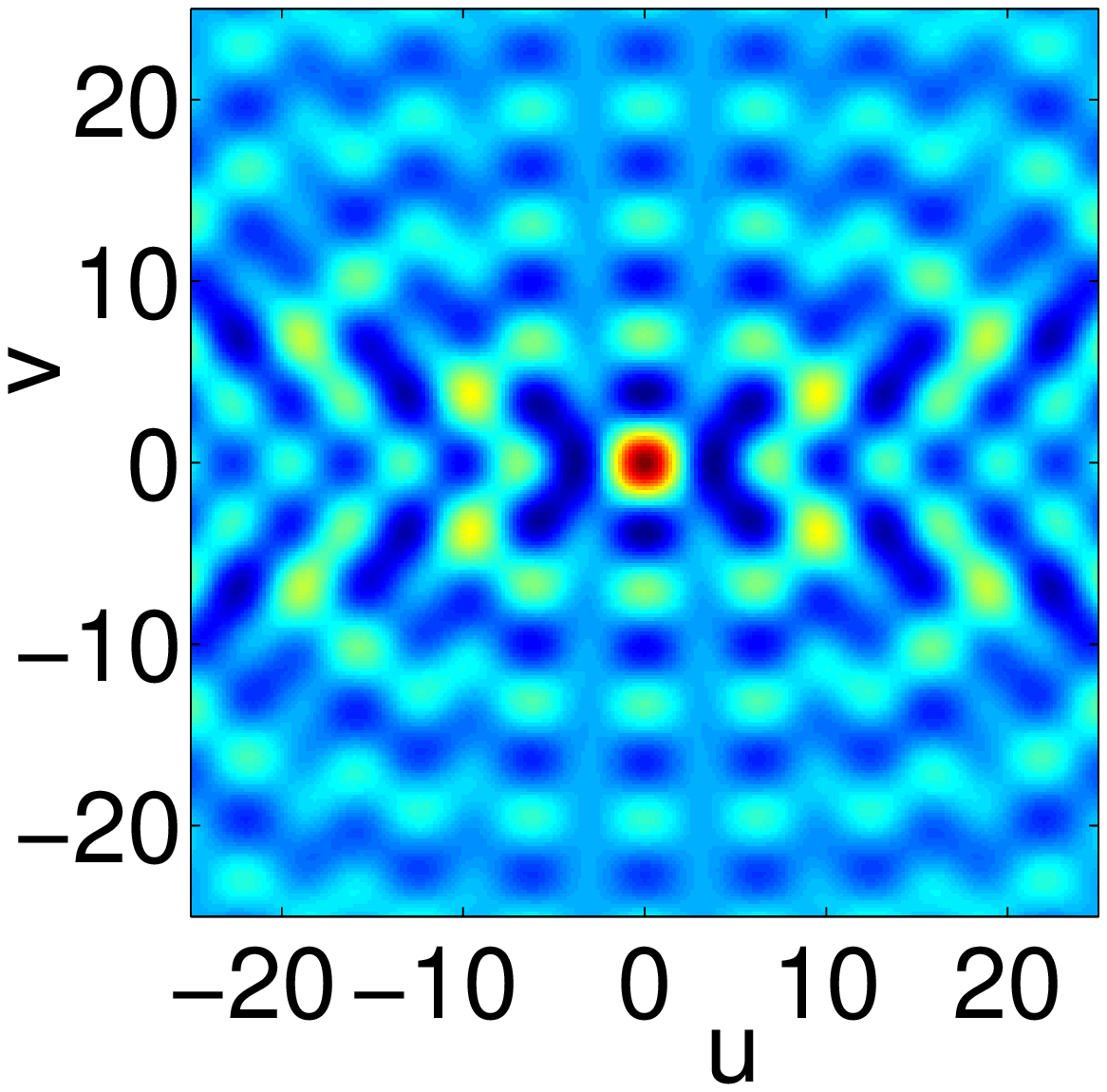}
\includegraphics[width=4.4cm,  angle=0]{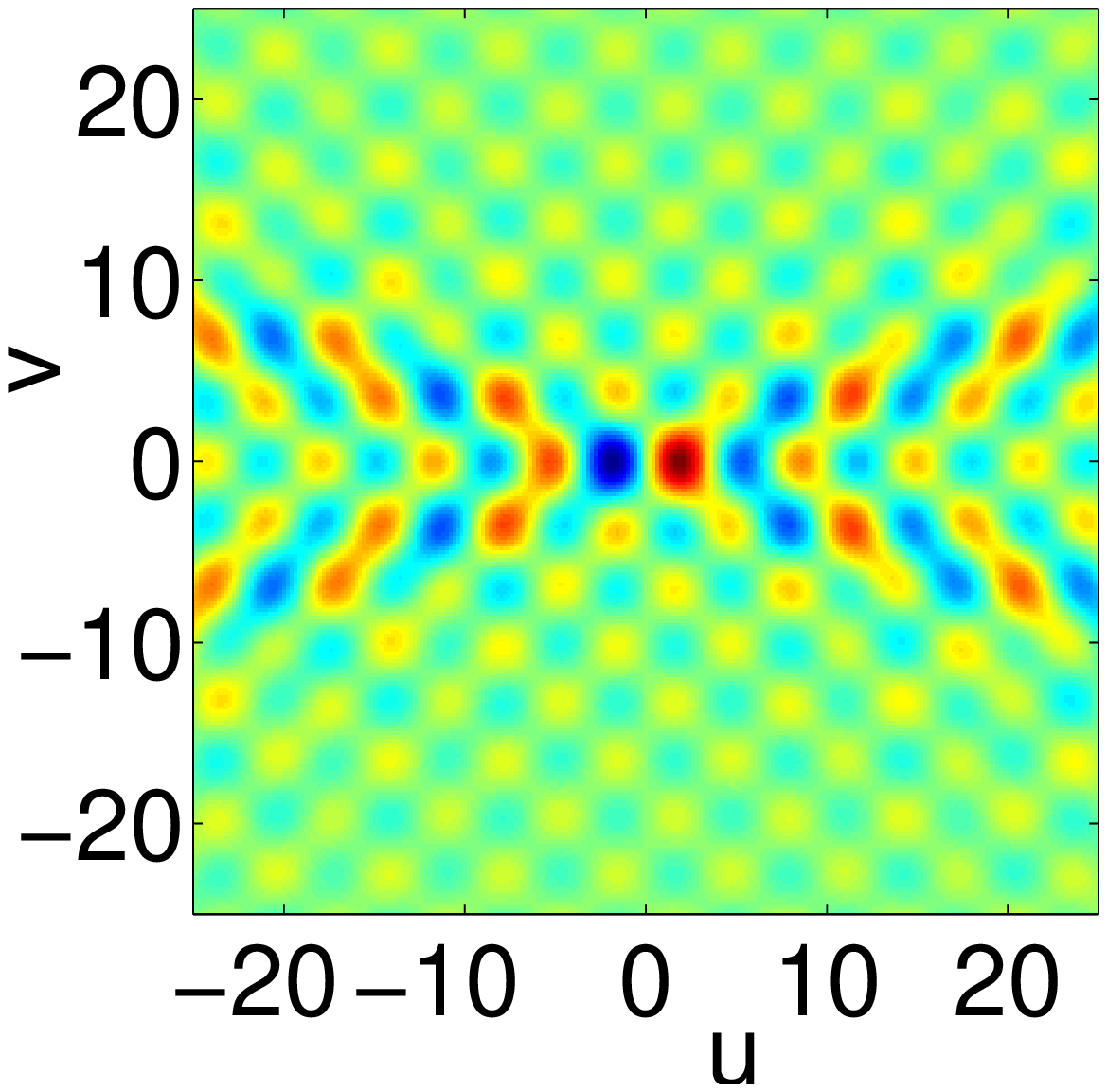}
\includegraphics[width=4.4cm,  angle=0]{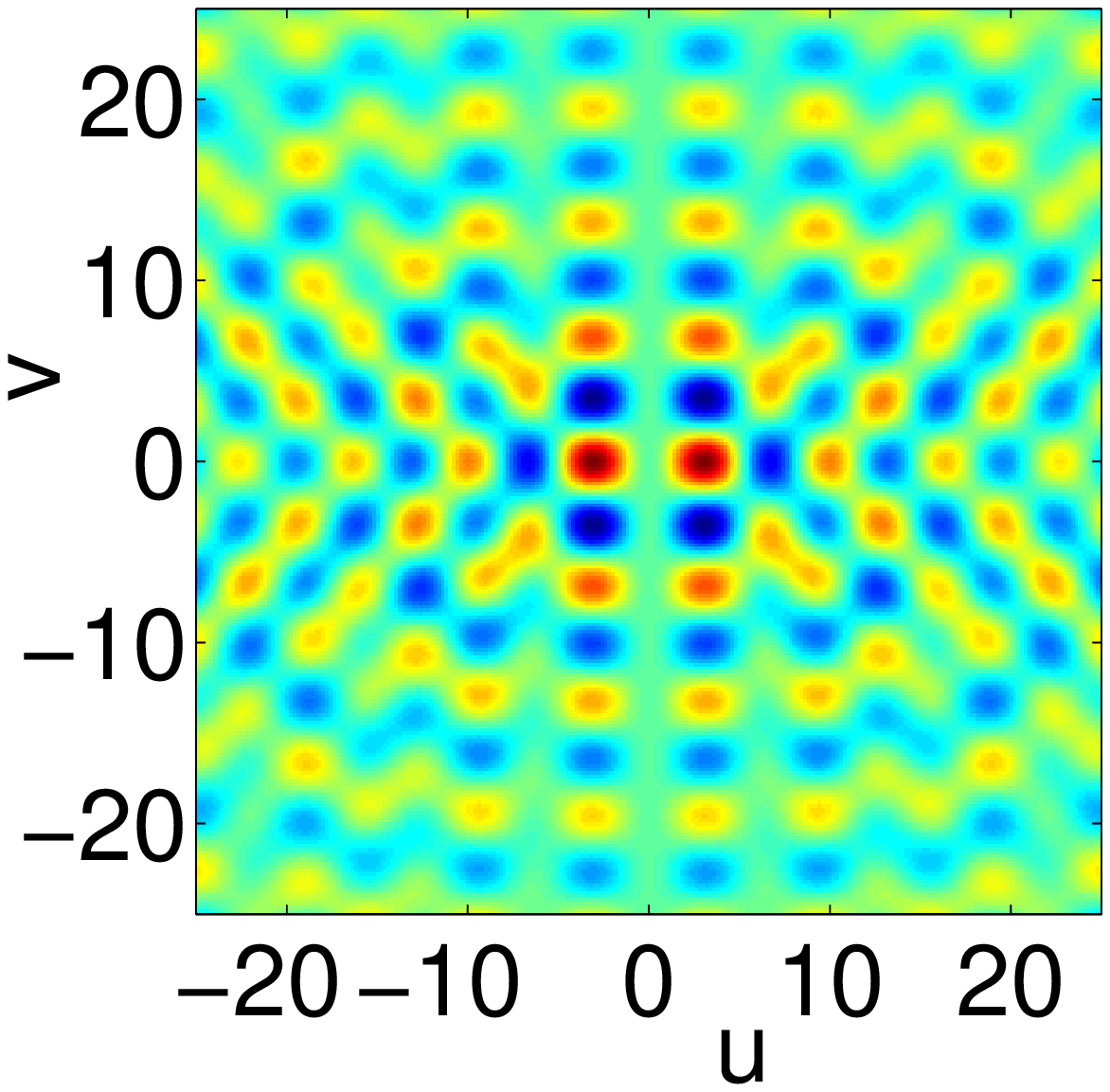}
\caption{\label{besselp_12_n}
Color map of the two-dimensional, one-parameter  Bessel function
$J_n^{1,2}(u,v;\ri)$ for $n=0$, $n=1$, $n=2$ (from
left to right). The figures show the real part.}
\end{figure}

Figure \ref{besselp_12_n} shows the real part of the Bessel functions 
$J_n^{1,2}(u,v;\ri)$
for $n=0,\,1,\,2$. 
A comparison with figure \ref{bessel_12_n} shows that the structure of the
functions is strongly altered by the angle parameter $\delta$.
\subsection{Differential equations and recurrence relations}
Finally, we will derive recurrence relations for the 
$J_{n}^{p,q}(u,v)$ and their derivatives. 
Differentiating the generating function (\ref{gen-funct-gen})
with respect to $u$ leads to
\begin{eqnarray}
&& {\textstyle \frac12}\big(z^p-z^{-p}\big)
\,\re^{\textstyle\frac{u}{2}(z^p-z^{-p})+\frac{v}{2}(z^q-z^{-q})} \nonumber \\
&&\quad ={\textstyle \frac12}\sum\limits_{n=-\infty}^\infty 
J_{n}^{p,q}(u,v)\,\big(z^{n+p}-z^{n-p}\big)
=\sum\limits_{n=-\infty}^\infty \partial_uJ_{n}^{p,q}(u,v)\,z^n\,.
\end{eqnarray}
Equating the coefficients of $z^n$, we find
\begin{eqnarray}
2 \,\partial_uJ_{n}^{p,q}(u,v)=J_{n-p}^{p,q}(u,v)-J_{n+p}^{p,q}(u,v)
\label{diffrule1}
\end{eqnarray}
and similarly
\begin{eqnarray}
2 \partial_vJ_{n}^{p,q}(u,v)=J_{n-q}^{p,q}(u,v)-J_{n+q}^{p,q}(u,v)\,.
\label{diffrule2}
\end{eqnarray}
If we differentiate (\ref{gen-funct-gen}) with respect to $z$ and
compare the coefficients, we find the recurrence equation
\begin{eqnarray}
pu\,\big(J_{n-p}^{p,q}(u,v)\!+\!J_{n+p}^{p,q}(u,v)\big)
\!+\!qv\,\big(J_{n-q}^{p,q}(u,v)\!+\!J_{n+q}^{p,q}(u,v)\big)
\!=\!2n J_{n}^{p,q}(u,v)\,.
\label{recurrence}
\end{eqnarray}
These are generalizations of the relations derived
by Reiss \cite{Reis80} for the case $J_{n}^{1,2}(u,v)$.

Similarly one can show that the derivatives of the generalized Bessel functions
(\ref{genp-bessel-sum-1q}) are given by
\begin{eqnarray}
  && 2 \partial_u J^{1,q}_{n}(u,v;\re^{\ri \delta}) =
  J^{1,q}_{n-1}(u,v;\re^{\ri \delta}) - J^{1,q}_{n+1}(u,v;\re^{\ri \delta}) \nonumber \\
  && 2 \partial_v J^{1,q}_{n}(u,v;\re^{\ri \delta}) =
  \re^{\ri \delta} J^{1,q}_{n-q}(u,v;\re^{\ri \delta}) -
  \re^{- \ri \delta} J^{1,q}_{n+q}(u,v;\re^{\ri \delta}).
\end{eqnarray}
Using these relations one can show that the two-dimensional Bessel functions
solve a variety of linear partial differential equations, depending on
their indices $(p,q)$.
These differential equation can be constructed systematically by adding up
derivatives of different order such that the different terms $~J^{p,q}_n$
cancel each other for every value of $n$.
For example the ordinary two-dimensional Bessel functions with $q=p$ solve the wave
equation
\begin{equation}
  \left( \partial_u^{2} - \partial_v^{2} \right) J_n^{1,1}(u,v) = 0,
  \label{eqn-wave-Bessel}
\end{equation}
while the generalized Bessel functions with $(p,q) = (1,2)$ and $\delta = \pi/2$
solve the time-dependent Schr\"odinger equation \cite{Datt91a,Datt92b}
\begin{equation}
  \ri \partial_v J^{1,2}_{n}(u,v;\ri) = \left( - 2 \partial_u^2 -1 \right) J^{1,2}_{n}(u,v;\ri).
  \label{eqn-Schroedinger-Bessel}
\end{equation}

Furthermore, a repeated application of the differentiation rules 
(\ref{diffrule1}), (\ref{diffrule2}) and
the recurrence relations (\ref{recurrence}) leads to the coupled differential equations
\begin{eqnarray}
   && \left[ \left( p u \partial_u + q v \partial_v \right)^2
   + \left( p^2 u \partial_u + q^2 v \partial_v \right)
   + p^2 u^2 + q^2 v^2 -n^2 \right] J^{p,q}_n(u,v) \nonumber \\
   && \quad = - p q u v \left( J^{p,q}_{n-p+q}(u,v) + J^{p,q}_{n+p-q}(u,v) -2 J^{p,q}_n(u,v) \right)
   \label{eqn-coupled-pde}
\end{eqnarray}
for arbitrary indices $(p,q)$.
Except from the right-hand side this equation is structurally similar to the
defining differential equation of the ordinary one-dimensional Bessel functions.
Equations (\ref{eqn-coupled-pde}) can be decoupled for $q = \nu p, \; \nu \in \mathbb{Z}$ by
applying again (\ref{recurrence}).
For $(p,q) = (1,\pm 1)$ this yields
\begin{equation}
   \left[ \left( u \partial_u + v \partial_v \right)^2
   + \left( u \partial_u + v \partial_v \right)
   +  (u \pm v)^2 - n^2 \right] J^{1,\pm 1}_n(u,v)
   = 0,
\end{equation}
while the respective calculations for other values of $q$ lead to
more complicated results.

\section{Polynomial expansion for small arguments}
We will first
analyze the regime of small arguments $u$ and $v$ and derive a leading
order polynomial expansion.
Following Wasiljeff \cite{Wasi69}, we expand  the $u,v$-dependent part of the
exponential function in (\ref{int-rep}) in a Taylor series:
\begin{eqnarray}
&&J^{p,q}_n(u,v)=\frac{1}{2\pi}\int\limits_{-\pi}^{+\pi}
\re^{\ri (u \sin{p t}+v\sin{qt}-n t)}\rd t\\
&&\quad =\frac{1}{2\pi}\sum\limits_{k=0}^\infty\frac{1}{2^kk!}\int\limits_{-\pi}^\pi
\big(u\re^{\ri pt}-u\re^{-\ri pt}+v\re^{\ri qt}-v\re^{-\ri qt}\big)^k\,
\re^{-\ri nt}\rd t\,.
\end{eqnarray}
Using the polynomial formula
\begin{eqnarray}
(a+b+c+d)^j=j!\sum\limits_{\alpha,\beta,\sigma ,\zeta}\nolimits^{\,'}
\,\frac{a^\alpha\,}{\alpha!}\frac{b^\beta}{\beta!}\frac{c^\sigma}{\sigma!}\frac{d^\zeta}{\zeta!}\,,
\end{eqnarray}
where the primed sum runs over all indices with $j=\alpha+\beta+\sigma+\zeta$,
one obtains after rearranging terms and carrying out the integration
the series expansion 
\begin{eqnarray}
J^{p,q}_n(u,v)=\sum\limits_{j=0}^\infty \frac{1}{2^j}
\sum\limits_{\alpha,\beta,\sigma ,\zeta}\nolimits^{\,''}
\frac{u^{\alpha+\beta}\,v^{\sigma +\zeta}}{\alpha !\,\beta !\,\sigma !\,\zeta !}\,.
\label{sum-dp}
\end{eqnarray}
Here the double-primed sum includes all nonnegative integers with
\begin{eqnarray}
j=\alpha+\beta+\sigma+\zeta\quad \textrm{and} \quad n=(\alpha-\beta)p+(\sigma-\zeta)q\,.
\label{small-cond}
\end{eqnarray}
The sum can be transformed into a more convenient form by
introducing $\ell =\alpha+\beta$, $2f =\ell+\alpha-\beta$,
$m=\sigma +\zeta$ and $2g=m+\sigma -\zeta$. After some elementary algebra,  this yields
\begin{eqnarray}
J^{p,q}_n(u,v)=\sum\limits_{\ell,m\ge 0}\,a_{\ell,m}^{(n,p,q)}\,\frac{u^m v^\ell}{2^{\ell +m}}
\end{eqnarray}
with
\begin{equation}
a_{\ell,m}^{(n,p,q)}=
\sum_{\substack{f=0,\ldots,+\ell \cr g=0,\ldots,+m \cr n=p(2f-\ell)+q(2g-m)}}
\frac{1}{f!(\ell-f)!g!(m-g)!}\,.
\end{equation}
The lowest order approximation in this expansion can be found in explicit
form for the case $p=1$. Then the lowest order term in (\ref{sum-dp}) is
given by\footnote{Notation: $\lfloor x \rfloor$ is the largest integer $\le x$,
$\lceil x \rceil$ is the smallest integer $>x$ and ${\rm mod}(n,q)=\frac{n}{q}-\lfloor \frac{n}{q}\rfloor$.}
\begin{eqnarray}
\qquad (\alpha,\,\beta,\,\sigma,\,\zeta)=\left\{
\begin{array}{ll} ( n- q\lfloor \frac{n}{q}\rfloor,\,0,\, \lfloor \frac{n}{q}\rfloor,\, 0)
&\textrm{if } \frac{n}{q} \notin \mathbb N \textrm{ and } 2q\,{\rm mod}(n,q)\le q+1\\[2mm]
( 0,\, -n+ q\lceil \frac{n}{q}\rceil,\, \lceil \frac{n}{q}\rceil, 0)\
&\textrm{if } \frac{n}{q}\notin \mathbb N \textrm{ and } 2q\,{\rm mod}(n,q)\ge q+1\\[2mm]
( 0,\,0,\,\frac{n}{q},\, 0) &\textrm{if } \frac{n}{q}\in \mathbb N \,.\hfill (\theequation)\stepcounter{equation}
\end{array}\right.\nonumber
\end{eqnarray}
In  most cases it is given by a single term, as for example in
\begin{eqnarray}
J_3^{1,2}(u,v)\sim \frac{1}{2^2}\,uv \quad ,\quad J_4^{1,2}(u,v)\sim \frac{1}{2^2}\,v^2\,.
\end{eqnarray}
Note that, because $n=3$ is not a multiple of $q=2$, the Bessel function
$J_3^{1,2}$ is identically equal to zero on the $v$-axis (see eq.~(\ref{symm-n=mq})), however
$J_3^{1,2}$ and $J_4^{1,2}$ do not vanish on the $u$-axis, where we have 
$J_n^{1,2}(u,0)=J_n(u)\sim (u/2)^n/n!$
(see eq.~(\ref{pq=1q-v=0})).

In certain cases more terms of the same minimum order $j$ appear.
This happens if $\frac{n}{q} \notin \mathbb N$ and  $2q\,{\rm mod}(n,q)= q+1$.
One can easily check that this requires that $q$ is odd, $q=2\nu+1$, and
$n=\mu q +\nu +1$ with $\nu,\mu \in \mathbb N$. In this case, the lowest
order approximation reads
\begin{eqnarray}
J^{1,q}_n(u,v)\sim 
\frac{u^\nu v^\mu}{2^{\nu+\mu+1}\nu !\mu !}\left( \frac{u}{\nu+1}+ \frac{v}{\mu+1}\right)\,.
\end{eqnarray}
This yields the (approximate) nodal line
\begin{eqnarray}
v=-\frac{\mu +1}{\nu +1}\,u 
\label{smalluv-nodes}
\end{eqnarray}
for small $u$ and $v$.
As an example, we note $q=5$ and $n=23$, i.e.~$\nu=2$ and $\mu=4$ and therefore
\begin{eqnarray}
J^{1,5}_{23}(u,v)\sim 
\frac{u^2 v^4}{2^{7}2!4!}\left( \frac{u}{3}+ \frac{v}{5}\right)\,.
\end{eqnarray}
\section{Asymptotic approximations}
\label{s-asymptotic}
In the examples of two-dimensional Bessel functions $J_n^{p,q}(u,v)$ shown in figures 
\ref{bessel_12_n} and \ref{bessel_13_n} for $(p,q)=(1,2)$ and $(p,q)=(1,3)$, respectively,
one observes a rich oscillatory structure which will be analyzed in the following.
The skeleton of this structure and valuable approximations can be obtained 
asymptotically by means of the  stationary phase approximation
\begin{eqnarray}
\int\limits_{-\pi}^\pi\rd t\,
 h(t)\,\re^{\ri \,g(t)}\simeq \sum_{t_\stat }{\textstyle \sqrt{\frac{2\pi}{\pm
 g''(t_\stat )}\,}}
\,h(t_\stat )\,\re^{\ri g(t_\stat )\pm \ri \pi/4}\,.
\label{statphase}
\end{eqnarray}
The sum extends over all contributing real stationary points and
the $\pm$ sign is chosen so that $\pm g''(t_\stat)$ is positive (see, e.g., \cite{Mars87} for more
details). Previous studies of asymptotic approximations for two-dimensional
Bessel functions \cite{Niki64,Leub81,Reis03} have been restricted
to the case $p=1$, $q=2$ and special regions of the index $n$ and arguments
$u,v$.

Information about the oscillatory structure of the multivariable Bessel functions
$J_n^{p,q}(u,v)$
can be obtained from asymptotic approximations for large arguments and/or
large indices. We will consider three of the large number of possible limits:
the case when both arguments $u$ and $v$ are large, whereas $n$ remains fixed,
and the case where one argument, $v$, and the index $n$ are large for
fixed value of the argument $u$. Finally we will consider
the limit where both variables as well as the index $n$ are large. 
In all cases we assume small fixed values of
the indices $p$ and $q$.
\subsection{Basic structure for large arguments $u$ and $v$}
\label{s-asymptotic1}
We base our analysis on the integral representation (\ref{int-rep})
\begin{eqnarray}
J_{n}^{p,q}(u,v)=\frac{1}{2\pi}
\int_{-\pi}^{+\pi}\rd t\,
\re^{\ri (\phi(t)-nt)}\ , \quad \phi(t)=u\sin pt+v\sin qt\,.
\label{int-rep-copy}
\end{eqnarray}
Here we will consider the asymptotic limit of large arguments $u$ and $v$ 
assuming that $n$ is fixed, i.e.~we
identify $g(t)=\phi(t)$ in an application of (\ref{statphase}). 
The condition
\begin{eqnarray}
\phi'(t)=pu\cos pt+qv\cos qt =0
\label{stat-phase}
\end{eqnarray}
determines the stationary points  $t_\stat $ (note that there are always pairs
of such stationary points with different sign due to the symmetry of the cosine-function).
The integral (\ref{int-rep-copy}) is then approximately given by
\begin{eqnarray}
J_{n}^{p,q}(u,v)=\sum_{t_\stat}  \frac{1}{\sqrt{2\pi |\phi''(t_\stat )|}}
\,\re^{\ri(\phi(t_j)-nt_\stat \pm \pi/4)}\,,
\label{int-stat-phase-gen}
\end{eqnarray}
where the $\pm$-sign is given by the sign of $\phi''(t_j)$. The main contribution
to the sum is provided by real-valued stationary points, complex points lead to
exponentially decaying terms.
At the points where two stationary points coalesce when the arguments $u$ and $v$ are
varied, the second derivative
\begin{eqnarray}
\phi''(t)=-p^2u\sin pt-q^2v\sin qt
\label{stat-phase-ss}
\end{eqnarray}
vanishes and the approximation diverges. Crossing these bifurcation points, the
function changes its character. In the present case, the 
bifurcations are determined by the simultaneous solution of
\begin{eqnarray}
pu\cos pt_\stat =-qv\cos qt_\stat   \quad \textrm{and}\quad
p^2u\sin pt_\stat =-q^2v\sin qt_\stat \,.
\label{stat-phase-bif}
\end{eqnarray}
This can be most easily satisfied if both sides of one of the two equations are
equal to zero. We distinguish two cases:\\[2mm]
Case (i) $\sin pt_\stat =-\sin qt_\stat =0$\,: For coprime $p$ and $q$, this implies
$t_\stat =0$ or $t_\stat=\pi$ and therefore (from  $pu\cos pt_\stat =-qv\cos qt_\stat $) we have
$pu=-qv$ or $pu=-(-1)^{p+q}\,qv$, respectively. We therefore obtain
the bifurcation lines
\begin{eqnarray}
v=\pm p\,u/q \quad &&\textrm{if one of the } p,q \textrm{ is even}\\[2mm]
v=-p\,u /q\quad &&\textrm{else }. 
\label{bif1}
\end{eqnarray}
Case (ii) $\cos pt_\stat =\cos qt_\stat =0$\,: This implies $pt_\stat =\pi/2+j\pi$ and $qt_\stat =\pi/2+k\pi$
with integer $j$ and $k$ or \,$(2k+1)p=(2j+1)q$\, and (for coprime $p$ and $q$)
\,$p=2j+1$\, and \,$q=2k+1$\,.  With \,$\sin pt_\stat =\sin  (\pi/2+j\pi)=(-1)^j$\, and 
\,$\sin qt_\stat =\sin  (\pi/2+k\pi)=(-1)^k$\,
the second condition in (\ref{stat-phase-bif})
leads to
\begin{eqnarray}
v=-(-1)^{j+k}\,u\,p^2/q^2 \quad && p=2j+1\ ,\ q=2k+1\,.
\label{bif2}
\end{eqnarray}

The examples in figure \ref{bessel_12_n} show Bessel functions $J_{n}^{1,2}(u,v)$
for various values of $n$.
Here $q$ is even and from equation
(\ref{bif1}) we find the bifurcation lines
\begin{eqnarray}
v=\pm u/2\,.
\label{bif-12}
\end{eqnarray}
We observe that the structure of the Bessel functions changes if one crosses these
lines. In the left and right sectors, we have only two stationary points, $\pm t_1$,
whereas in the upper and lower sectors we have four stationary points,  $\pm t_1$
and $\pm t_2$, and consequently a richer interference pattern. This will be analyzed
in more detail below.

For the two-dimensional Bessel function $J_n^{1,3}(u,v)$ displayed in
fig.~\ref{bessel_13_n} both upper indices are odd
and the bifurcation lines are given by eqs.~(\ref{bif1}) and (\ref{bif2}):
\begin{eqnarray}
v=- \,u/3 \quad \textrm{and}\quad v=u/9 \,.
\label{bif-13}
\end{eqnarray}
The qualitative difference to the behavior of $J_n^{1,2}(u,v)$
in fig.~\ref{bessel_12_n} is obvious.

Let us now analyze the function \,$J_n^{1,2}(u,v)$\, in more detail working
out explicitly the stationary phase approximation.
In view of the symmetry $J_n^{1,2}(u,v)=J_{-n}^{1,2}(-u,-v)$ (\ref{properties1})
we can assume $v>0$ in the following for simplicity.
The stationary phase condition
\begin{eqnarray}
u\cos t=-2v\,\cos 2t=-2v\,(2\cos^2t-1)
\end{eqnarray}
can be solved for $c=\cos t$ with solution
\begin{eqnarray}
c_{\pm} =\frac18\,\big(-u/v\pm \sqrt{(u/v)^2+32\,}\,\big)\,.
\end{eqnarray}
In the region $-2<u/v<+\infty$ the necessary condition $|c_\pm|\le 1$ 
is met by $c_+$ and vice versa by $c_-$ in the region
$-\infty <u/v<+2$. Note that in the interval $-2<u/v<+2$ {\it both\/}
solutions fulfill $|c_\pm|\le1$.
With  \,$t_\pm=\arccos c_\pm$\, and \,$\sin t_\pm=\pm\sqrt{1-c^2_\pm}$\, 
we arrive at
\begin{eqnarray}
\phi_\pm&=&u\sin t_\pm+v\sin 2t_\pm= \pm(u+2vc_\pm)\,\sqrt{1-c^2_\pm} \\
\phi''_\pm&=&-u\sin t_\pm-4v\sin 2t_\pm= \mp(u+8vc_\pm)\,\sqrt{1-c^2_\pm}
\end{eqnarray}
and with the definitions
\begin{eqnarray}
F_+(u,v)\!\!&=&\!\!\left\{\begin{array}{l l}
\sqrt{\frac{2}{\pi |\phi''_+ |}}\,\cos \big(\phi_+ -n\arccos c_+ -\frac{\pi}{4}\big)
 &\textrm{for } -2v<u\\[2mm] 0 & \textrm{else }\end{array}\right.\\
F_-(u,v)\!\!&=&\!\!\left\{\begin{array}{l l}
\sqrt{\frac{2}{\pi |\phi''_-|}}\,\cos \big(\phi_- +n\arccos c_- -\frac{\pi}{4}\big)
 &\textrm{for } u<+2v\\[2mm] 0 & \textrm{else }\end{array}\right.\,\\
\label{F+-}\nonumber
\end{eqnarray}
the final result can be written as
\begin{eqnarray}
J_{n}^{1,2}(u,v)\simeq F_+(u,v)+F_-(u,v)\,.
\label{J12-semi}
\end{eqnarray}
Here one should be aware of the fact that in the region \,$|u|\ll 2|v|$\, {\it both}
of the terms $F_\pm(u,v)$ provide a non-vanishing contribution.

\begin{figure}[t]
\begin{center}
\includegraphics*[width=6.5cm]{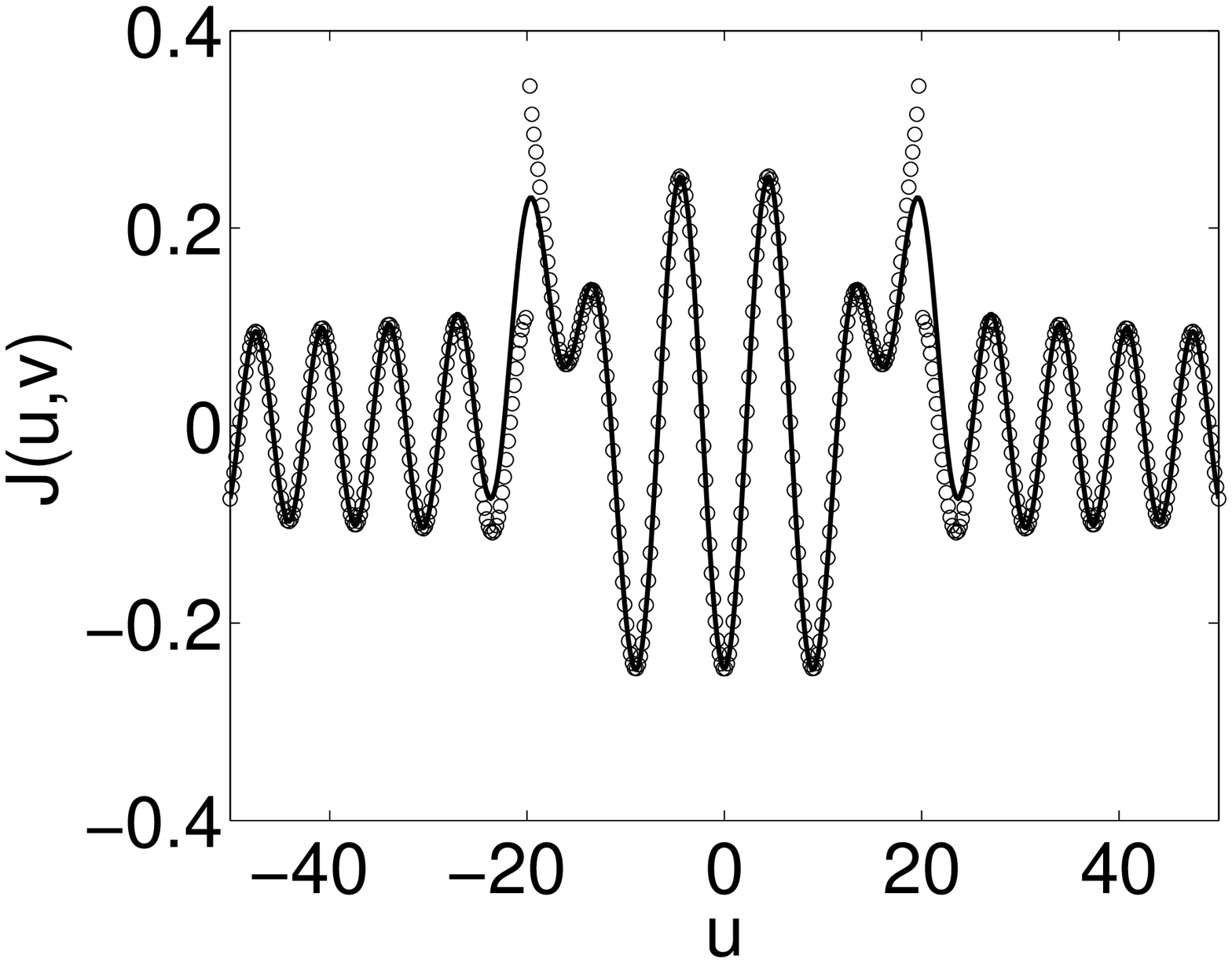}
\includegraphics*[width=6.5cm]{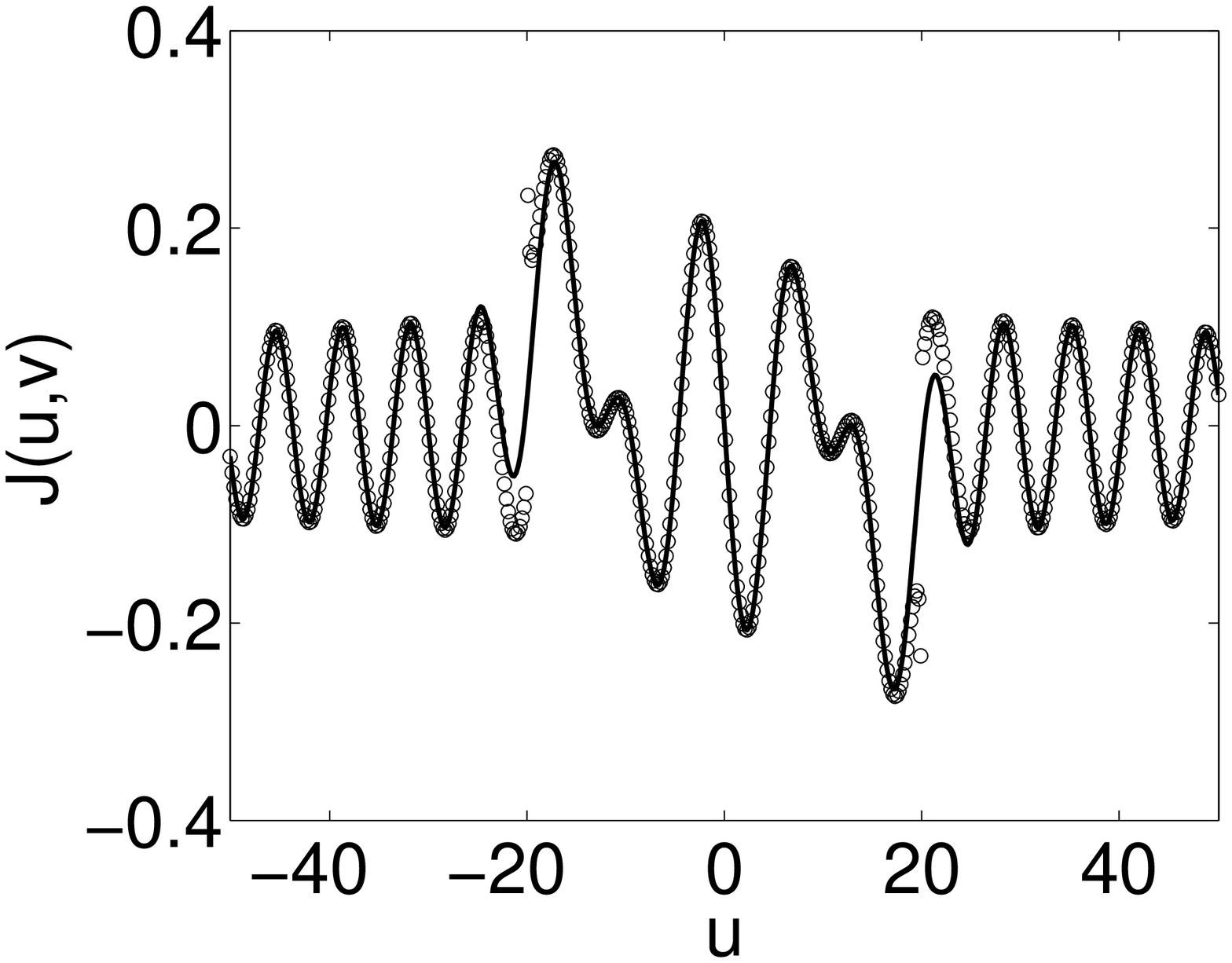}
\end{center}
\caption{\label{semi12} Two-dimensional Bessel functions $J_n^{1,2}(u,v)$
with $v=10$ for  $n=0$ (left) and $n=1$ (right) as a function of $u$
(full curve) in comparison with the stationary phase approximation 
(\ref{J12-semi}) (open circles).
}
\end{figure}

This so-called 'primitive' stationary phase approximation diverges
at the bifurcation lines $v=\pm 2u$. If desired, it can be improved by
taking complex stationary points into account and by taming the divergences
by uniformization methods.

In the limit $|u|\rightarrow \infty$ the asymptotic approximation (\ref{J12-semi})
simplifies drastically. Only $F_+$ contributes for $u>0$ ($F_-$ for 
$u<0$) and with $t_\pm=\pm \pi/2$, $\phi_\pm=\phi''_\pm=\pm u$
we find
\begin{eqnarray}
J_{n}^{1,2}(u,v)\simeq
\sqrt{\frac{2}{\pi |u|}}\,\cos \big(u-n\frac{\pi}{2} -\frac{\pi}{4}\big)\,,
\label{J12-semi-v=0}
\end{eqnarray}
which agrees with the well-known asymptotic approximation of
the ordinary Bessel function $J_n(u)$ for large arguments \cite{Abra72}.
In the alternative limit  $v\rightarrow \infty$ both terms, $F_+$ and $F_-$ 
contribute. With
$c_\pm=-\frac{u}{8v}\pm \frac{1}{\sqrt{2}}$, $\phi_\pm=v\pm\frac{u}{\sqrt{2}}$
and $\phi''_\pm=\pm 4v$  we obtain
\begin{eqnarray}
J_{n}^{1,2}(u,v)\simeq
\sqrt{\frac{2}{\pi v}}\left\{\begin{array}{ll}
+\cos \big(v-(n+1)\frac{\pi}{4}\big)\,\cos \frac{u}{\sqrt{2}}\ &n \textrm{ even}\\[2mm]
-\sin \big(v-(n+1)\frac{\pi}{4}\big)\,\sin \frac{u}{\sqrt{2}}\ &n \textrm{ odd}\,,
\end{array}\right.
\end{eqnarray}
as already derived in \cite{Datt96}.

As an illustration of the asymptotic formula (\ref{J12-semi}), fig.~\ref{semi12} shows the two-dimensional
Bessel function  $J_n^{1,2}(u,v)$ in comparison with the stationary phase approximation
for $v=10$ and $n=0$, $1$.
With the exception of the vicinity of the divergences at $u=\pm 2v=\pm 20$, the
agreement is excellent. 
This approximation can be used in oder to determine
the nodal lines of the two-dimensional Bessel functions
which is of interest for applications in physics \cite{05bicro}.
\subsection{Large argument $v$ and large index $n$}
\label{s-asymptotic2}
In this section we will consider the regime  where the index $n$ and 
one of the arguments, e.g.~$v$, are large. In view of 
(\ref{properties1}), we can assume $n\ge 0$.
Following the analysis applied to the
special case  $J^{1,2}_n(u,v)$ by Reiss and Krainov \cite{Reis03}, we 
separate the integral representation  (\ref{int-rep}) into a fast, 
$\re^{\ri g(t)}$, and
a slowly oscillating part:
\begin{eqnarray}
J^{p,q}_n(u,v)=\frac{1}{2\pi}\int\limits_{-\pi}^\pi\rd t\,
 \re^{\ri u\sin{pt}}\re^{\ri g(t)}
\quad ,\quad g(t)=v\sin{qt}-nt\,,
\label{IntB}
\end{eqnarray}
and evaluate the integral approximately by the method of stationary phase or
the saddle point method if the stationary points are complex valued
(for details see, e.g., \cite{Mars87}).

The stationary
points  $t_\stat $ of  $g(t)$ are obtained from $g'(t_\stat) =0$ as
\begin{eqnarray}
\cos{qt_\stat }=\frac{n}{qv}
\end{eqnarray}
with real valued solutions for $n<q|v|$ and complex solutions
otherwise. We discuss these two cases separately.\\

\noindent
(i) For   $n<q|v|$  the stationary points are
\begin{eqnarray}
t_\stat^\pm =\pm (t_0+2\pi \stat/q)\ ,\quad \stat=0,\,1,\, 2,\,\ldots 
\ ,\quad t_0=\frac{1}{q}\,\arccos \frac{n}{qv} \,,
\label{saddle1}
\end{eqnarray}
i.e.~a finite number in the interval $-\pi < t_\stat^\pm \le \pi$.

With $\sin qt_\stat^\pm =\pm \sin qt_0$ we have
\begin{eqnarray}
g(t_\stat^\pm)&=&v\sin qt_\stat^\pm -nt_\stat^\pm =\pm v\sin t_0 \mp
n(t_0+2\pi\stat/q)\,]\\[1ex]
g''(t_\stat^\pm )&=&-q^2v\sin qt_\stat =\mp q^2v \sin qt_0 \,.
\end{eqnarray}
and, using \,$\sin pt_\stat^\pm=\pm \sin p(t_0+2\pi\stat/q)$, the final result is
\begin{eqnarray}
&J_n^{p,q}(u,v)\simeq 
{\textstyle \sqrt{\frac{1}{2\pi q^2v \sin qt_0}\,}} \sum\limits_{\stat ,\pm}
\,\re^{\pm \ri \,[\, u\sin p(t_0+\frac{2\pi\stat}{q}) + v\sin qt_0 - n(t_0+\frac{2\pi \stat}{q})
-\frac{\pi}{4}\,]}\nonumber\\
&={\textstyle \sqrt{\frac{2}{\pi q^2v \sin qt_0}}}\sum\limits_{\stat}
\cos \big[ u\sin p(t_0+{\textstyle\frac{2\pi\stat}{q}}) + v\sin qt_0 
- n(t_0+{\textstyle\frac{2\pi \stat}{q}})-
{\textstyle\frac{\pi}{4}}\big].
\label{Jpq-semi1}
\end{eqnarray}
We will work out the case $p=1$ and $q=2$ in more
detail. Here we find four stationary points $\pm (t_0+\stat \pi)$
with $\stat =0$ and $-1$ and therefore
\begin{eqnarray}
&&J_n^{1,2}(u,v)\simeq
{\textstyle \sqrt{\frac{1}{2\pi v \sin 2t_0}}}
\,\Big\{\cos\, \big[\,u\sin t_0 +v\sin 2t_0 - nt_0-\frac{\pi}{4}\,\big]\nonumber \\
&&\qquad \qquad \qquad +(-1)^n\cos\, \big[-\,u\sin t_0 +v\sin 2t_0 
- nt_0-\frac{\pi}{4}\,\big]\Big\}
\label{J12-semi1}\\[1ex]
&&={\textstyle \sqrt{\frac{2}{\pi v \sin 2t_0}}}
\,\left\{\begin{array}{ll}
+\cos (u\sin t_0)\,\cos (v\sin 2t_0 -nt_0-\frac{\pi}{4})\ & n \textrm{ even} \\[1ex]
-\sin (u\sin t_0)\,\sin (v\sin 2t_0 -nt_0-\frac{\pi}{4})&  n \textrm{ odd}
\end{array}
\right.\nonumber 
\end{eqnarray}
with 
\begin{eqnarray}
t_0=\frac12 \,\arccos \frac{n}{2v}\ , \quad
\sin 2t_0=\sqrt{1-\frac{n^2}{4v^2}\,}\ , \quad 
\sin t_0=\sqrt{\frac12-\frac{n}{4v}\,}\,.
\end{eqnarray}
In comparison with the semiclassical approximation derived in section
\ref{s-asymptotic1}, the result (\ref{J12-semi1}) agrees approximately with
(\ref{J12-semi}) also for small values of $n$, as for example $J_0^{1,2}(u,v)$ shown
in figure \ref{semi12} for $v=10$. Equation (\ref{J12-semi1}) misses however the
structural transition at $|u|=2v$ and cannot describe the region $|u|>2v$.\\

\noindent
(ii) For   $n>q|v|$  the stationary points (\ref{saddle1}) are complex,
$t_\stat=x_\stat+\ri y$  with real part 
\begin{eqnarray}
x_\stat =
\left\{ \begin{array}{ll}2s\pi/q &,\ v>0\\
(2s+1)\pi/q&,\ v<0\end{array}\right.
 \ ,\ \stat =0,\,\pm 1,\,\pm 2,\ldots 
\label{saddle2x}
\end{eqnarray}
with $-\pi < x_\stat \le +\pi$. The imaginary part is the same for all $\stat$:
\begin{eqnarray}
y_\pm =\pm \frac{1}{q}{\rm arccosh}{\left(\frac{n}{q|v|}\right)}\,.
\label{saddle2y}
\end{eqnarray}
The integral is approximately carried out
by the saddle point integration, where the integration path is deformed
to a steepest decent curve passing through the saddle points \cite{Mars87}: 
\begin{eqnarray}
\int\limits_{-\pi}^\pi\rd t\,
 h(t)\,\re^{\ri g(t)}\simeq \sum_{s}{\textstyle \sqrt{\frac{2\pi}{-
 \ri g''(t_\stat)}\,}}
\,h(t_\stat)\,\re^{\ri g(t_\stat )}\,.
\end{eqnarray}
The second derivative is
\begin{eqnarray}
\ri g''(t_\stat)=-\ri v q^2\sin{qt_\stat}=-q^2v\sinh{qy}
\end{eqnarray}
and the  conditions for the integration path \cite{Mars87} can only be 
satisfied for 
the saddle points in the upper (lower) complex plane for $v>0$ ($v<0$). 
With
\begin{eqnarray}
\ri g(t_\stat)=\ri(v\sin qt_\stat-nt_\stat)=-v\sinh qy\,-\ri n x_\stat-ny
\end{eqnarray}
we obtain the result
\begin{eqnarray}
J_{n}^{p,q}(u,v)=\frac{\re^{-|v\sinh qy|-n|y|}}{\sqrt{2\pi q^2|v\sinh qy|}}\,
\sum\limits_\stat\re^{-\ri nx_\stat}\re^{\ri u\sin{p(x_\stat+iy)}}.\label{approx}
\label{Jpq-semi2}
\end{eqnarray}
Let us again consider the special case $p=1$, $q=2$ in more detail. Two saddle
points contribute 
($x_\stat=\pm \pi/2$ with $y_-$ for $v<0$ or  
$x_\stat=0$ and $\pi$  with $y_+$ for $v>0$) and equation (\ref{Jpq-semi2}) 
simplifies.\\[2mm] 
For $v<0$ we find
\begin{eqnarray}
J_{n}^{1,2}(u,v)=\frac{\re^{+v\sinh 2y-ny}}
{\sqrt{-2\pi v \sinh 2y}}
\ \cos{\big(u \,\cosh y\,-n\,\frac{\pi}{2}\big)}
\label{J12-semi2m}
\end{eqnarray}
with
\begin{eqnarray}
y=\frac12 \,{\rm arccosh} \frac{n}{2|v|}\ , \quad
\sinh 2y=\sqrt{\frac{n^2}{4v^2}-1\,}\ , \quad 
\cosh y=\sqrt{\frac{n}{4|v|}+\frac12\,}
\end{eqnarray}
in agreement with the result derived in \cite{Reis03}.\\[2mm]
For $v>0$ the resulting approximation is non-oscillatory:
\begin{eqnarray}
J_{n}^{1,2}(u,v)=\frac{\re^{-v\sinh 2y-ny}}
{\sqrt{4\pi{\sqrt{2v \sinh 2y}}}}
\left\{ \begin{array}{ll}
\cosh (u\sinh y) & n \textrm{ even}\\[1ex]
\sinh (u\sinh y) & n \textrm{ odd}
\end{array}\right.
\label{J12-semi2p}
\end{eqnarray}
with
\begin{eqnarray}
y=\frac12 \,{\rm arccosh} \frac{n}{2v}\ ,\quad  \sinh y=\sqrt{\frac{n}{4v}-\frac12\,}\,.
\end{eqnarray}
Note that both asymptotic approximations (\ref{J12-semi1}) and (\ref{J12-semi2p})
satisfy the symmetry relation $J_n(-u,v)=(-1)^n\,J_n(u.v)$ (cf.~eq.~(\ref{symm-q-even})\,).

\begin{figure}[t]
\begin{center}
\includegraphics*[width=0.30\textwidth]{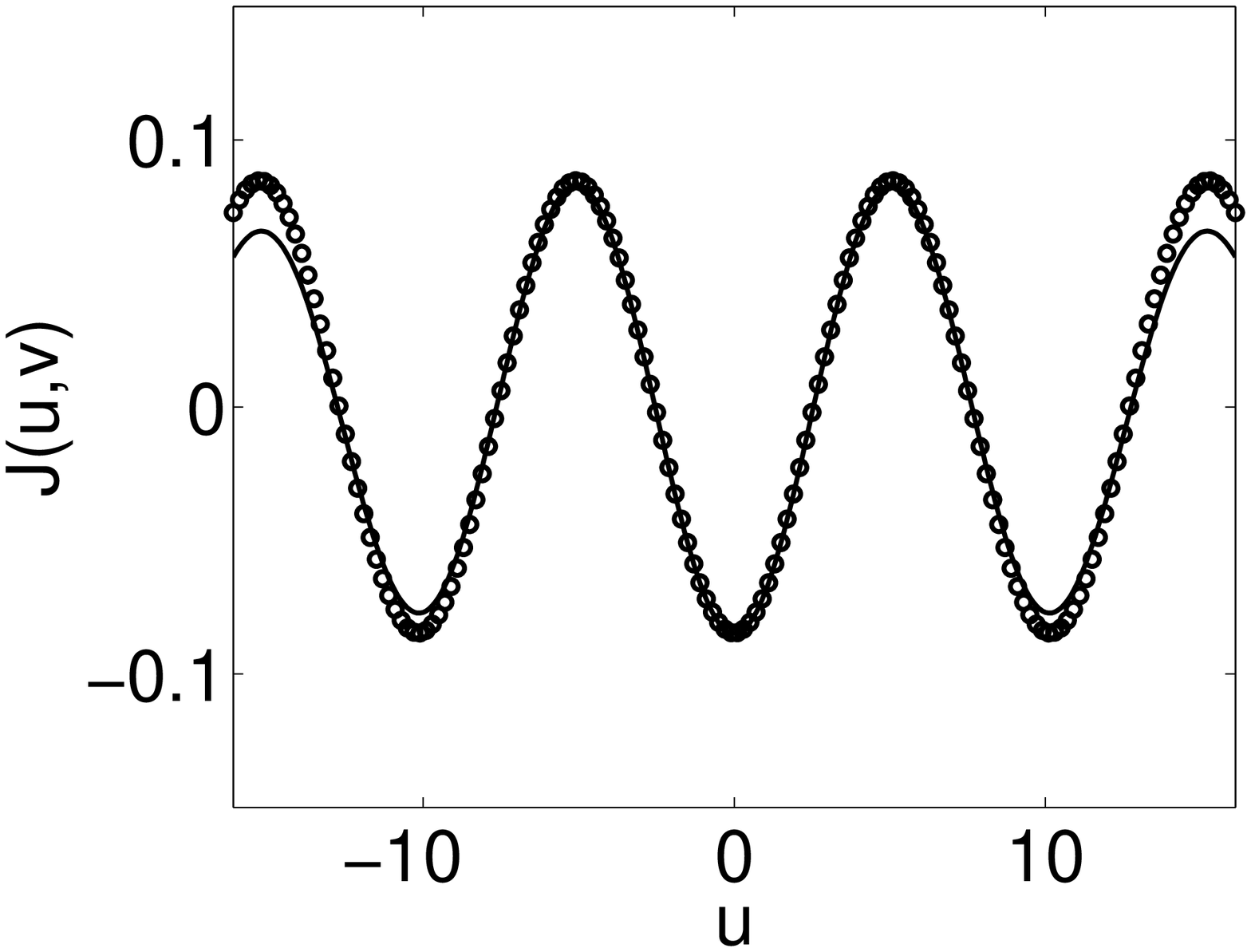}
\hspace{1mm}
\includegraphics*[width=0.30\textwidth]{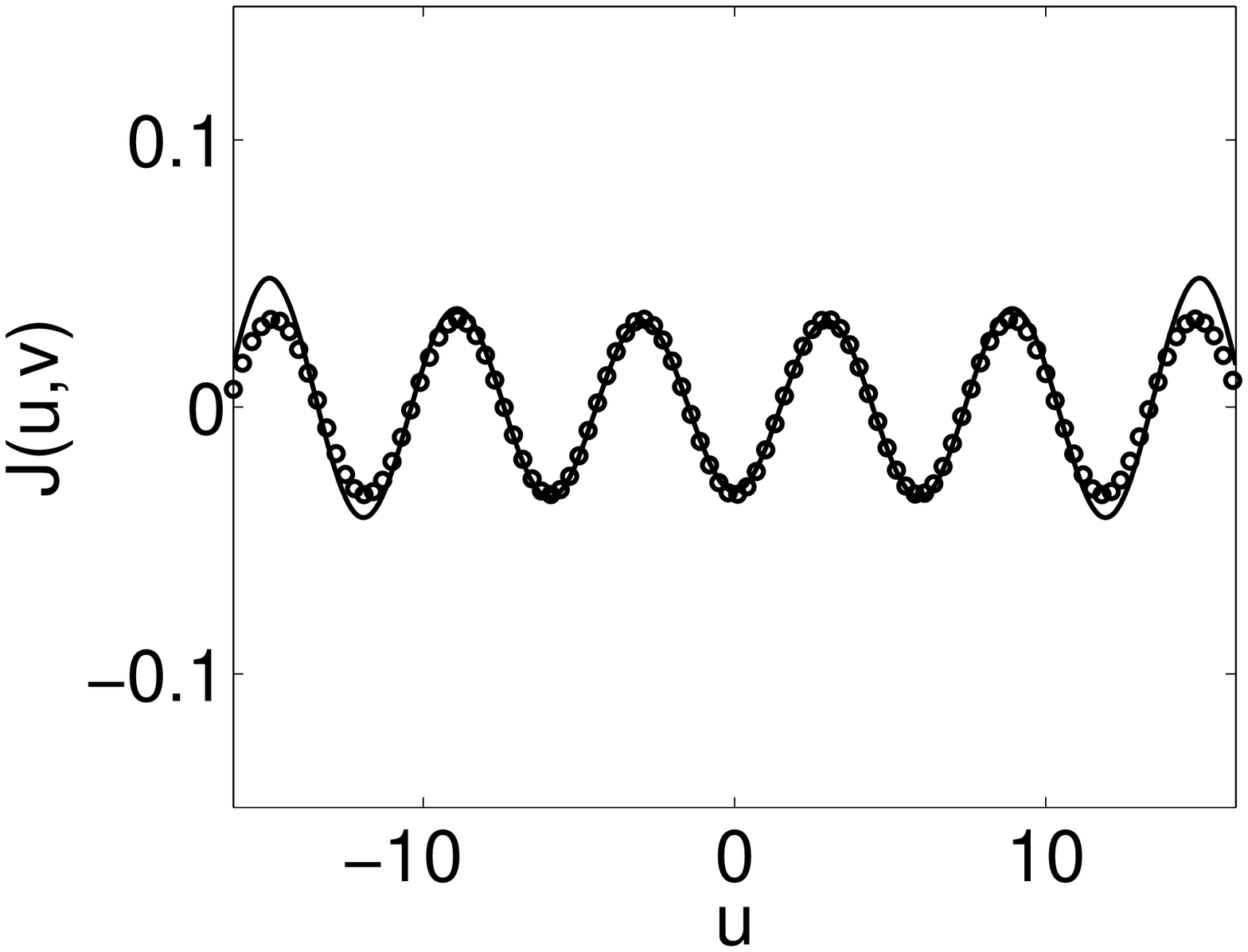}
\hspace{1mm}
\includegraphics*[width=0.30\textwidth]{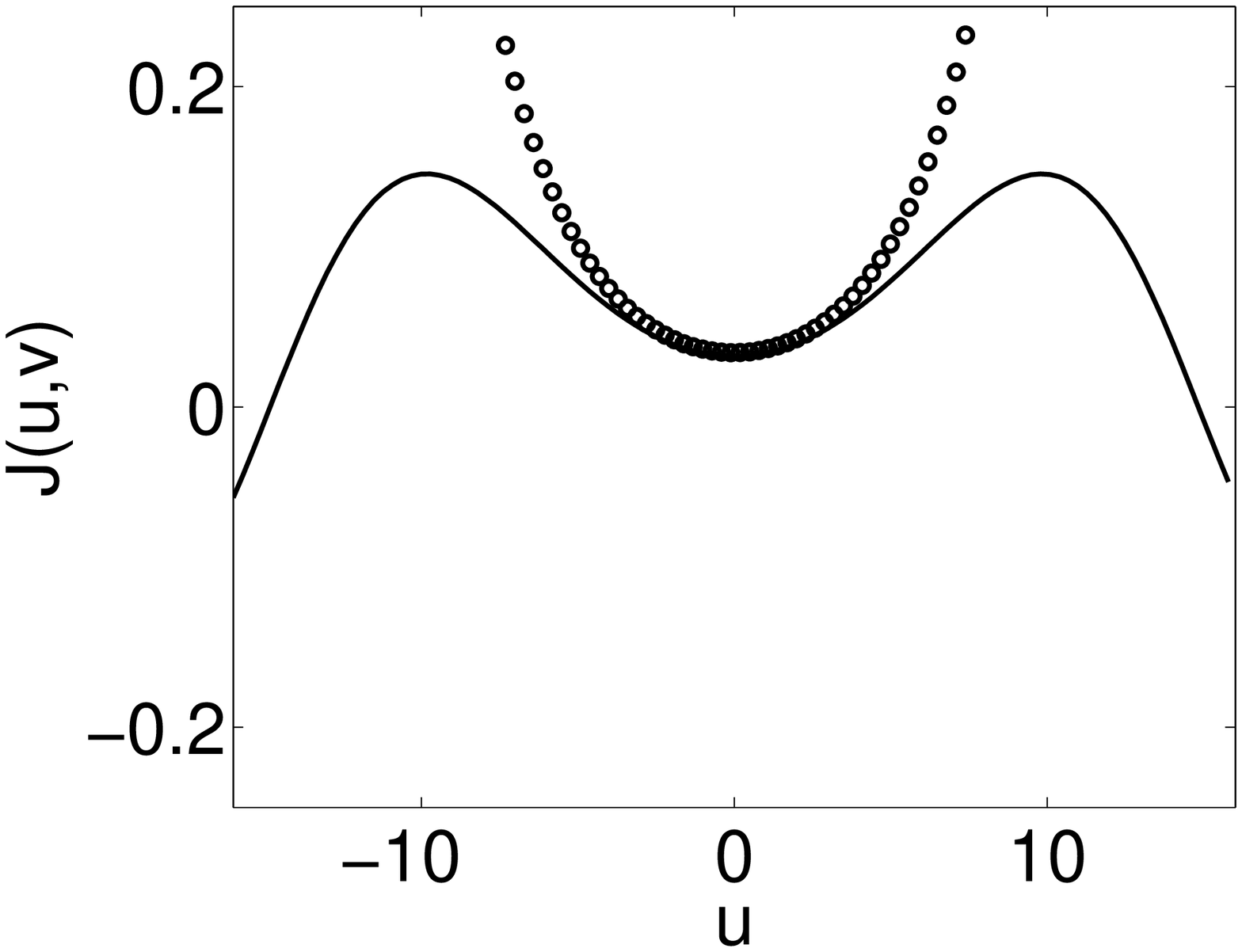}
\end{center}
\caption{\label{AbbStatio}
Comparison of the asymptotic approximation (\ref{J12-semi2m}) (open circles) with the exact
two-dimensional Bessel function $J_{n}^{1,2}(u,v)$ (full line) for $n=30$.
Left: Case (i) for $v=64$ using eq.~(\ref{J12-semi1}). Middle: Case (ii) with
$v=-12$ using eq. (\ref{J12-semi2m}).
Right: Case (ii) with $v=+12$ using eq. (\ref{J12-semi2p})}
\end{figure}

Figure \ref{AbbStatio} demonstrates the quality of the asymptotic approximation
for $n=30$  and $v=64$ (case (i)) or   $v=-12$ (case (ii)). Reasonable
agreement is observed 
for $|u|<10$. These simple approximations get worse in the vicinity of
$n=q\,|v|$ where they diverge. A finite result can be obtained using an
appropriate
uniformization technique, in the present case an Bessel uniformization, e.g.~a
mapping onto an (ordinary) Bessel function \cite{Stin73} (see also \cite{Leub81}
for an alternative method).

The asymptotic approximations (\ref{J12-semi1}) and (\ref{J12-semi2m}) provide explicit
estimates for the nodal lines of $J_{n}^{1,2}(u,v)$.
For $n<2|v|$ we find
\begin{eqnarray}
u\sqrt{\frac12-\frac{n}{4v}}=\left\{
\begin{array}{ll}
(2j+1)\frac{\pi}{2} \quad & n \textrm{   even}\\[2mm]
j\pi \quad & n \textrm{   odd}\end{array}\right.
\quad ,\quad j=0,\,\pm 1,\,\pm 2,\ldots
\end{eqnarray}
and for $n>2|v|$ we have for $v<0$ zeros at
\begin{eqnarray}
u\,\sqrt{\frac12-\frac{n}{4v}}=\big(2j+n\big)\frac{\pi}{2}\quad ,\quad j=0,\,\pm 1,\,\pm 2,\ldots\,.
\end{eqnarray}
These results are, of course, in agreement with the zeros observed in
fig.~\ref{AbbStatio}.
\subsection{Large arguments $u$, $v$ and large index $n$}
\label{s-large-uvn}

\begin{figure}[t]
\center
\includegraphics[width=6.0cm,  angle=0]{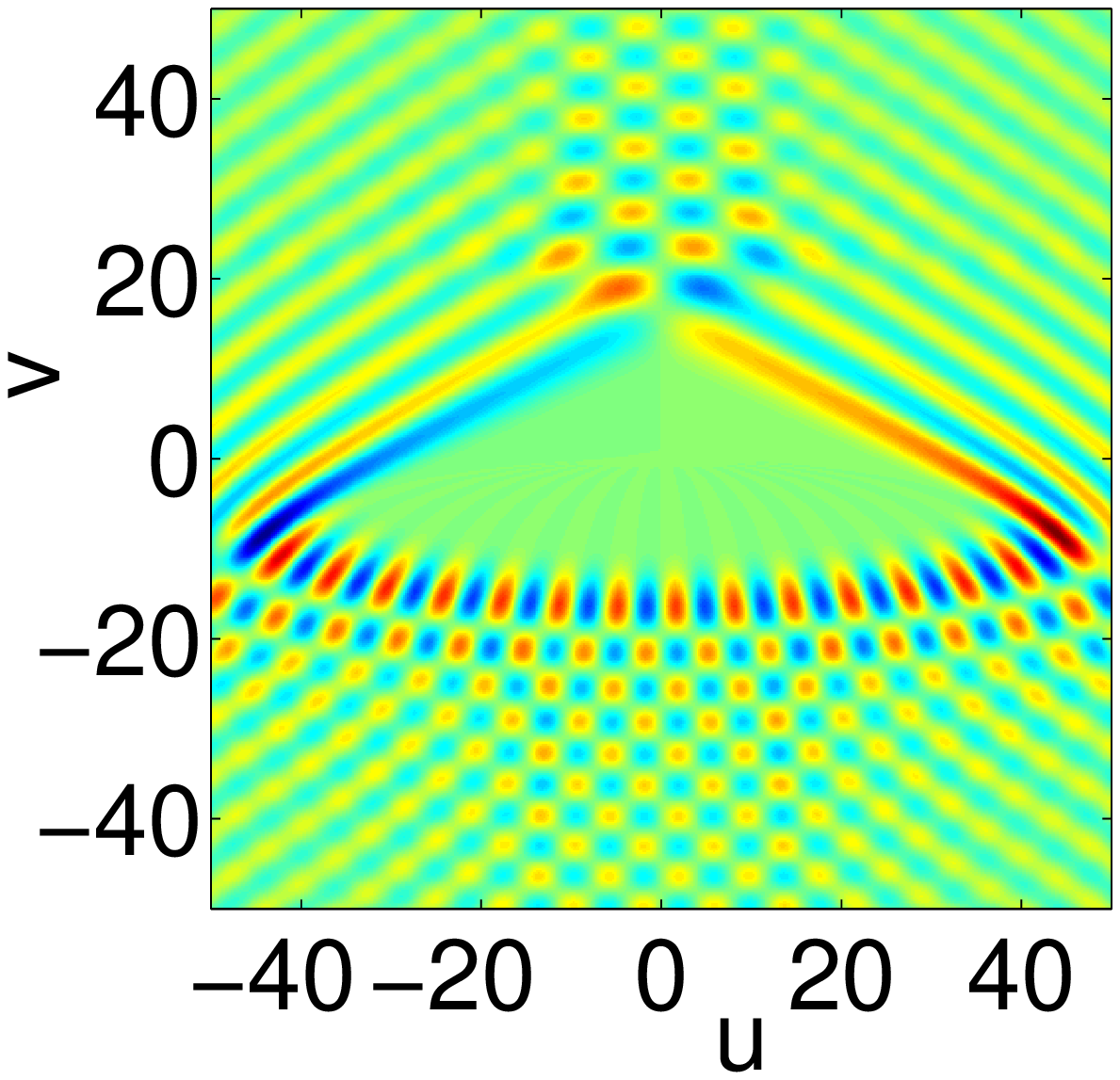}
\hspace{2mm}
\includegraphics[width=6.0cm,  angle=0]{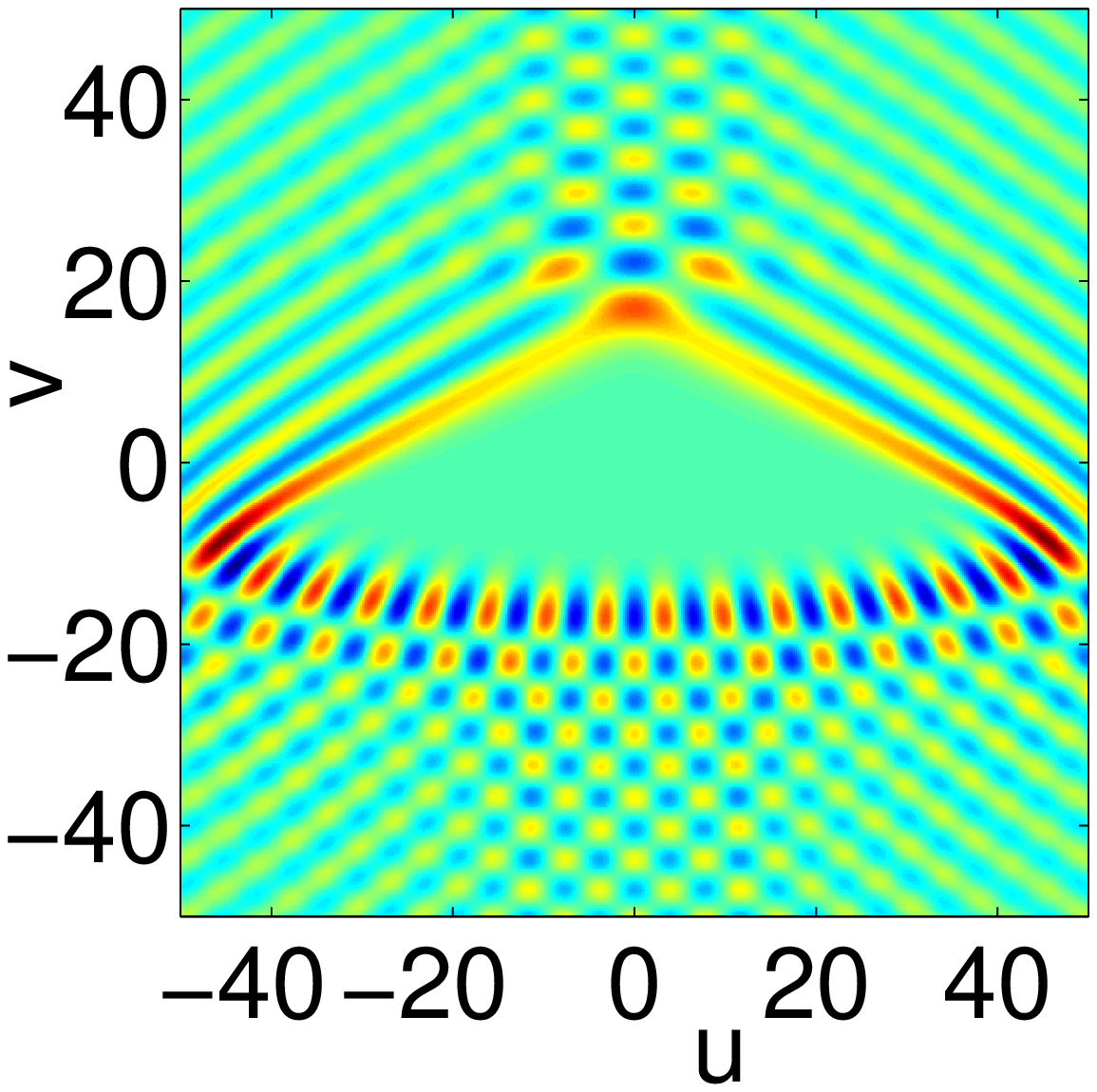}
\caption{\label{bessel_29+30_1_2}
Color map of the two-dimensional Bessel functions $J_n^{1,2}(u,v)$ for $n=29$ 
(left) and $n=30$ (right). Note the different symmetries of these functions. Their
overall structure can be explained by means of the bifurcation
set shown in fig.~\ref{bifurcation}.}
\end{figure}

As an example of the structure of the two-dimensional Bessel functions for
large indices,
fig.~\ref{bessel_29+30_1_2} shows  $J_n^{1,2}(u,v)$ for $n=29$ and $n=30$.
These functions look quite similar, they are clearly distinguished, however,
by their symmetry property $J_n^{1,2}(-u,v)=(-1)^nJ_n^{1,2}(u,v)$ 
(see eq.~(\ref{symm-q-even})), i.e. $J_{30}^{1,2}$ is even and  $J_{29}^{1,2}$
is odd with respect to a reflection $u \rightarrow -u$. Therefore  $J_{29}^{1,2}$
vanishes on the $v$-axis, $J_{29}^{1,2}(0,v)=0$. The function $J_{30}^{1,2}$
is symmetric on the $v$-axis:  $J_n^{1,2}(0,-v)=J_n^{1,2}(0,v)$ 
(see eq.~(\ref{symm-n=mq})),
despite of the apparent asymmetry with respect to the reflection 
$v \rightarrow -v$.

In additions to the oscillatory pattern in the four sectors,
we observe 
a region close to the center where the values of the
Bessel functions are small. This pattern
can again be explained by a consideration of the asymptotic limit
where both arguments and the index $n$ are large using 
\begin{eqnarray}
g(t)=u\sin pt+v\sin qt-nt
\end{eqnarray}
in the stationary phase approximation (\ref{statphase}). The stationary points
$t_\stat$ are determined by
\begin{eqnarray}
g'(t_\stat)=pu\cos pt_\stat +qv\cos qt_\stat -n =0\,.
\end{eqnarray}
The zeros of the second derivative
\begin{eqnarray}
g''(t)=-p^2u\sin pt-q^2v\sin qt
\end{eqnarray}
appearing in the denominator of (\ref{statphase}) determine 
the bifurcation set of these solutions.

\begin{figure}[t]
\center
\includegraphics[width=6.0cm,  angle=0]{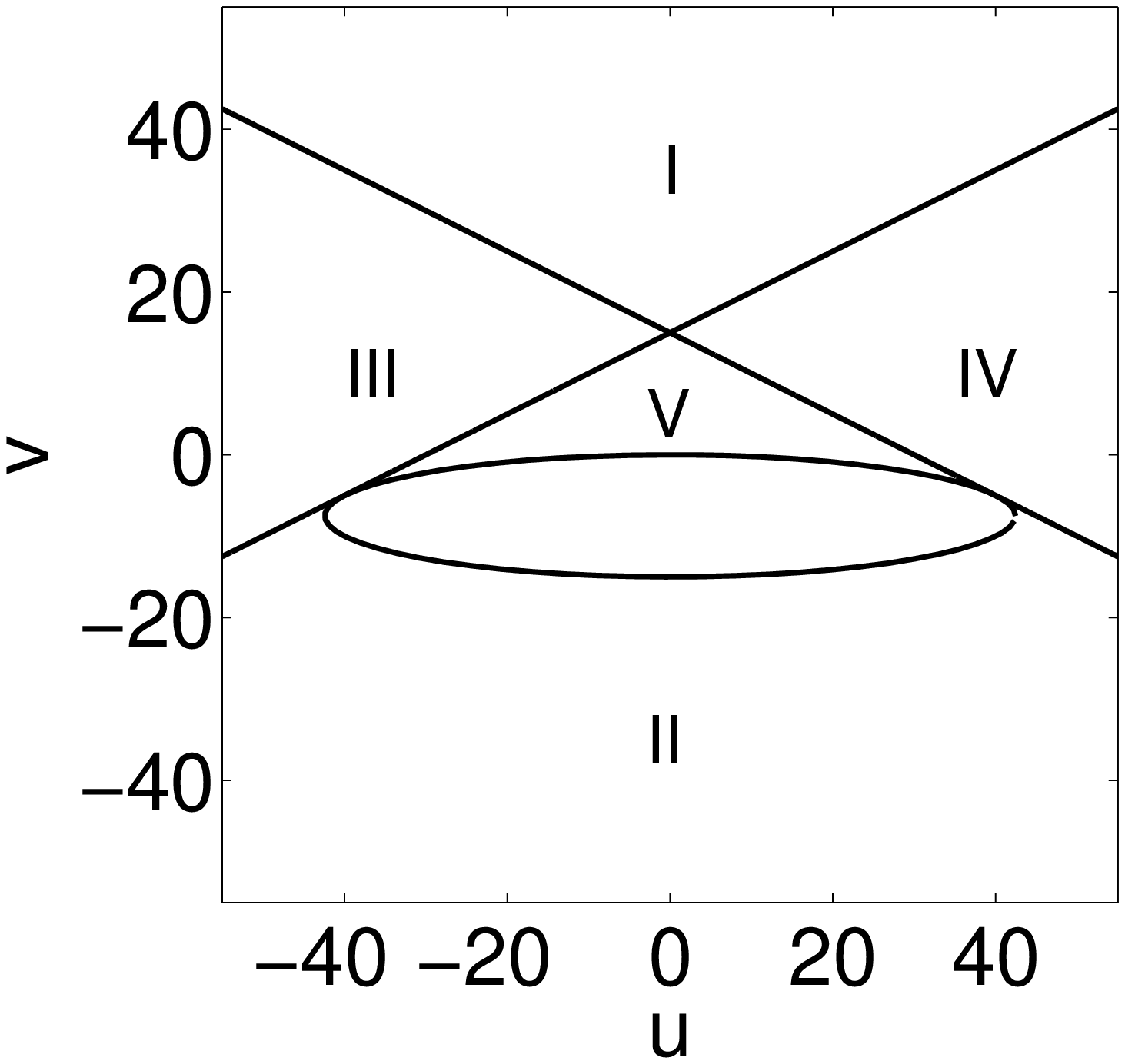}
\caption{\label{bifurcation}
Bifurcation curves of the stationary points for the
two-dimensional Bessel function $J_{30}^{1,2}(u,v)$.}
\end{figure}

Restricting ourselves again to the case $(p,q)=(1,2)$ these
equations simplify and can be solved in closed form:
\begin{eqnarray}
g'(t_\stat)&=&u\cos t_\stat +2v\cos 2t_\stat -n =0
\label{stat12}
\end{eqnarray}
with solutions
\begin{eqnarray}
c_\pm=\cos t_\stat = -\frac{u}{8v}\pm \sqrt{\Big(\frac{u}{8v}\Big)^2
+\frac12+\frac{n}{4v}\,}
\label{cpm}
\end{eqnarray}
(here again each solution $c_\pm$ implies two stationary points 
$t_\stat$ because of the symmetry of the cosine function).
The bifurcation set -- the skeleton of the Bessel function -- is found when
\begin{eqnarray}
g''(t_\stat)=-u\sin t_\stat -4v\sin 2t_\stat =0
\end{eqnarray}
is satisfied in addition to (\ref{stat12}).
Eliminating $t_\stat$ we find the
solutions
\begin{eqnarray}
v=(n\pm u)/2
\label{lines}
\end{eqnarray}
(note that  for $|u|\gg n$ these straight lines agree with the ones stated above 
in eq.~(\ref{bif-12}))
and the ellipse
\begin{eqnarray}
\frac{16(v+n/4)^2}{n^2}+\frac{u^2}{2n^2}=1
\label{ellipse}
\end{eqnarray}
centered at \,$(u,v)=(0,-n/4)$\, with half axes $\sqrt{2}\,n$ and $n/4$. Inside
this ellipse the stationary points (\ref{cpm}) are complex, outside they are
real. A brief calculation furthermore shows that the straight lines (\ref{lines})
are tangential to the ellipse (\ref{ellipse}). 
This bifurcation set is shown in fig.~\ref{bifurcation}. In the
upper sector (I) between the bifurcation lines we have $-1<c_\pm<+1$, as well
as in the lower sector (II) outside the ellipse. Hence we have four real solutions
$t_\stat$ in these regions and a corresponding oscillatory pattern. In
the right sector (IV) two of these solutions become complex because of
$|c_+|>1$ and similarly in the left hand sector (III) with  $|c_-|>1$.
In the triangular segment (V) in the lower sector above the ellipse, we have
$|c_+|>1$ and $|c_-|>1$ and therefore no real stationary points.
In the elliptic region with complex valued stationary points, the Bessel function is
damped but still oscillatory. An example is shown in fig.~\ref{AbbStatio}
(right hand side) which shows a cut through the Bessel function $J_{30}^{1,2}(u,v)$
shown in fig.~\ref{bessel_29+30_1_2} for $v=-12$ close to the elliptic bifurcation
curve. A cut at $v=64$ (left hand side) shows the oscillations in region (I).
Note that the semiclassical approximations shown in fig.~\ref{AbbStatio}
are the simplified versions developed in section \ref{s-asymptotic2}. A more
refined semiclassical treatment along the lines discussed above will provide a
much better agreement for larger values of $u$ (compare also the treatment in
\cite{Reis80}).

\section*{Acknowledgments}
Support from the Deutsche Forschungsgemeinschaft
via the Graduiertenkolleg  ``Nichtlineare Optik und Ultrakurzzeitphysik''
as well as from the ``Studienstiftung des deutschen Volkes''
is gratefully acknowledged.

\end{document}